\documentclass{elsarticle}

\usepackage{hyperref}
\usepackage{lscape}
\usepackage[dvipsnames]{xcolor}
\usepackage{amsmath}

\usepackage{etoolbox}
\usepackage{xspace}
\usepackage[english]{babel}
\usepackage{multirow}
\usepackage{subcaption}

\usepackage{ulem}

\journal{Space Science \& Technology}

\begin{document}
\begin{frontmatter}

\title{Machine learning technique for morphological classification of galaxies from the SDSS. \\III. Image-based inference of detailed features}

% \tnotetext[mytitlenote]{Fully documented templates are available in the elsarticle package on \href{http://www.ctan.org/tex-archive/macros/latex/contrib/elsarticle}{CTAN}.}

%% Group authors per affiliation:
%% or include affiliations in footnotes
%\author[mymainaddress,mysecondaryaddress]{Elsevier Inc}
%\ead[url]{www.elsevier.com}
%\author[mysecondaryaddress]{Global Customer Service\corref{mycorrespondingauthor}}
%\cortext[mycorrespondingauthor]{Corresponding author}
%\ead{support@elsevier.com}
%\address[mymainaddress]{1600 John F Kennedy Boulevard, Philadelphia}
%\address[mysecondaryaddress]{360 Park Avenue South, New York}

\author[mymainaddress]{V. Khramtsov\corref{mycorrespondingauthor}}
\cortext[mycorrespondingauthor]{Corresponding author}
\ead{vld.khramtsov@gmail.com}
\author[mysecondaryaddress]{I.B. Vavilova}
\author[mysecondaryaddress]{D.V. Dobrycheva}
\author[mysecondaryaddress]{M.Yu. Vasylenko}
\author[mysecondaryaddress]{\\O.V. Melnyk}
\author[mysecondaryaddress]{A.A. Elyiv}
\author[mymainaddress]{V.S. Akhmetov}
\author[mymainaddress]{A.M. Dmytrenko}
% \author[mymainaddress]{S.V. Khlamov}

\address[mymainaddress]{Institute of Astronomy, V.N. Karazin Kharkiv National University, 4 Svoboda Square, Kharkiv, 61000 Ukraine}
\address[mysecondaryaddress]{Main Astronomical Observatory of the NAS of Ukraine, 27, Akademik Zabolotny Str., Kyiv, 03143 Ukraine}

\begin{abstract}

This paper follows series of our works on the applicability of various machine learning methods to the morphological galaxy classification (Vavilova et al., 2021, 2022. We exploited the sample of $\sim 315\,800$ low-redshift SDSS DR9 galaxies with absolute stellar magnitudes of $-24^{m}<M_{r}<-19.4^{m}$ at $0.003<z<0.1$ redshifts as a target data set for the CNN classifier. Because it is tightly overlapped with the Galaxy Zoo 2 (GZ2) sample, we use these annotated data as the training data set to classify galaxies into 34 detailed features.   

In the presence of a pronounced difference of visual parameters between galaxies from the GZ2 training data set and galaxies without known morphological parameters, we applied novel procedures, which allowed us for the first time to get rid of this difference for smaller and fainter SDSS galaxies with $m_{r}<17.7$. We describe in detail the adversarial validation technique as well as how we managed the optimal train-test split of galaxies from the training data set to verify our CNN model based on the DenseNet-201 realistically. We have also found optimal galaxy image transformations, which help increase the classifier's generalization ability. 

We demonstrate for the first time that implication of the CNN model with train-test split of data sets and size-changing function simulating a decrease in magnitude and size (data augmentation) significantly improves the classification of smaller and fainter SDSS galaxies. It can be considered as another way to improve the human bias for those galaxy images that had a poor vote classification in the GZ project. Such an approach, likely auto-immunization, when the CNN classifier trained on very good images is able to retrain bad images from the same homogeneous sample, can be considered co-planar to other methods of combating the human bias.

The most promising result is related to the CNN prediction probability in classification of detailed features. The accuracy of CNN classifier is in the range of 83.3–99.4\,\% depending on 32 features (exception is for ``disturbed" (68.55\,\%) and ``arms winding medium" (77.39\,\%) features). As a result, for the first time, we assigned the detailed morphological classification for more than 140\,000 low-redshift galaxies, especially at the fainter end. A visual inspection of the samples of galaxies with certain morphological features allowed to reveal typical problem points of galaxy image classification by shape and features from the astronomical point of view. 

The morphological catalogs of low-redshift SDSS galaxies with the most interesting features are available through the UkrVO web-site \url{http://ukr-vo.org/starcats/galaxies/} and VizieR.
\end{abstract}

\begin{keyword}
galaxies: general — galaxies: fundamental parameters (classification) — methods: data analysis — techniques: image processing 
\end{keyword}

\end{frontmatter}

%\linenumbers

\section{Introduction}\label{sec:intro}

Convolutional neural network (CNN) as a machine learning (ML) technique is becoming more and more applicable for astronomical tasks. Its success has been proven sufficiently for big data observational sky surveys: galaxy classification by various properties, pattern recognition image description, celestial body's peculiarities identification, anomalies, transient object detection, etc. The CNNs are very helpful for finding and discovering previously unknown gravitationally lensed quasars \citep{Agnello2015, Ostrovski2017, Lanusse2018}, identifying gravitational lenses \citep{Jacobs2019, Khramtsov2019b, Petrillo2019, Ribli2019}, galaxy-galaxy strong gravitational lenses \citep{Pourrahmani2018} including in the Dark Energy Survey (DES) imaging data \citep{Pasquet2019} and weak gravitational lensing analysis to create galaxy images as an input \citep{Fussell2019}. The distance moduli estimates benefit from the CNNs utilization in the big data sets, which provide a wide number of galaxy features for learning (see review by Salvato et al. \citep{Salvato2019}). Bonnett et al. \cite{Bonnett2016} adopted the multiple ML methods for determining photometric redshifts with implications for weak lensing from the DES catalogue. Amaro et al. \cite{Amaro2019} compared ANNz2 \citep{Sadeh2016}, Bayesian photometric redshift method, and METAPHOR (Machine-learning Estimation Tool for Accurate PHOtometric Redshifts) for KiDS-ESO-DR3 and GAMA DR2 surveys. Similarly, Pasquet et al. \cite{Pasquet2018} used deep learning (DL) for classifying, detecting, and predicting photometric redshifts of quasars in the SDSS. ML and generative adversarial networks (GAN) were used to assign and predict photometric/spectroscopic redshifts within large-scale galaxy surveys with good accuracy \cite{Kugler2016, Speagle2017, Disanto2018, Salvato2019, Pasquet2019, Elyiv2020, Rastegarnia2022}. The ML approach serves as a basis for restoring galaxy distribution in the Zone of Avoidance \citep{Schawinski2017, Vavilova2018} and generating dark matter structures in cosmological models \citep{Diakogiannis2019, Tsizh2020, Chen2020}, for extraction information from noisy maps \cite{Moriwaki2021} and image reconstruction in the whole \citep{Flamary2016, Kremer2017}, for the task of deblending overlaps between foreground and background galaxies with GAN as CNN technique \citep{Reiman2019, Buchanan2021} (see, also, scalable ML algorithms and frameworks in \cite{Bouchefry2020}). The review on recent trends of ML applicability in cosmology and gravitational wave astronomy can be found in work by Burgazli et al. \cite{Burgazli2022}. 

The CNN models have expanded the ``optical" range of applications becoming useful for multiwavelength sky surveys. Among recent studies are as follows: search for blazar candidates in the Fermi-LAT Clean Sample \citep{Kang2019}; boosted decision tree for detecting the faint $\gamma$-ray sources with future Cherenkov Telescope Array \citep{Krause2017, Ruhe2020}; infrared colour selection of Wolf-Rayet candidates in our Galaxy using the Spitzer GLIMPSE catalogue \citep{Morello2018}; cosmic string searches in 21-cm temperature CMB maps \citep{Ciuca2017}; neural network-based Faranoff-Riley classifications of radio galaxies from the Very Large Array archive \citep{Aniyan2017} and DL classification of compact and extended radio source from Radio Galaxy Zoo \citep{Lukic2018}; CNN for morphological assignment to radio-detected galaxies with active nuclei \citep{Ma2019}. Scaife et al. \cite{Scaife2021} presented the first application of group-equivariant CNNs to radio galaxy classification and explored their potential for reducing intra-class variability by preserving equivariance for the Euclidean group on image translations, rotations, and reflections.

The merging galaxies are among the objects to be missclassified. Finding comprehensive samples of such galaxies in different merger stages is significant for studying these long-term processes. In this context, the adversarial training with Domain Adversarial Neural Networks (DANNs) altogether with the Maximum Mean Discrepancy (MMD) method was proposed by Ciprijanovic et al. \cite{Ciprijanovic2021}. Such adaptation techniques allowed these authors to demonstrate a great promise to classify galaxy mergers across domains. As well, to identify peculiar galaxies, an ML system needs to identify forms of galaxies that are not present in the dataset. For such identification of outlier galaxies, the unsupervised ML is proposed by Shamir et al. \citep{Shamir2021}.

Our work follows the previous paper by \cite{Vavilova2021a} (Paper I below), where the photometry-based approach for a binary morphological classification was applied to the SDSS DR9 set of low-redshift $\sim$ 315\,800 galaxies. Using various galaxy classification techniques (human labeling, multi-photometry diagrams, and five supervised ML methods), we found that the Support Vector Machine gives the highest accuracy (96.1\,\% early $E$ and 96.9\,\% late $L$ types). Determining the ability of each method to predict the galaxy morphological type, we verified various dependencies of the method's accuracy on redshifts, celestial coordinates, human labeling bias, the overlap of different morphological features, etc. 

The aim of this paper is to obtain the image-based classification of 315\,782 galaxies with absolute stellar magnitudes of $-24^{m}<M_{r}<-19.4^{m}$ at $0.003<z<0.1$ redshifts (with velocities correction on the velocity of Local Group, $V_{LG}>1500$ km/s). For this, we exploited the annotated data of the Galaxy Zoo 2 (GZ2) project with their crowd-sourcing strategy for volunteers to classify images by answering a series of questions. The sample of the GZ2 galaxies, which overlap with the studied galaxies, was served as the training data set for the CNN classifier. 

As compare to the paper by \cite{Vavilova2022} (Paper II below), this work investigates the problem of differences in the data sets in detail and suggest ways to overcome adversarial validation. We also use a neural network to predict some structural, morphological features that can help to classify galaxies with ware used by Walmsley et al. \cite{Walmsley2020}. We analyze the obtained samples of galaxies with different morphological features to obtain their quantitative/qualitative properties and to estimate an efficiency of CNN classifier.

We describe briefly the target, training, and inference galaxy data sets in Section 2. Methodology consisting of the data preparation, adversarial validation, CNN morphological classification with intelligent train-test split via adversarial scores is given in  Section 3 (see, also, Paper II). The general results and discussion are in Section 4, and the conclusion is presented in Section 5. 

\section{Galaxy data sets}\label{sec:data}

\subsection{Target data set}

One of the most crucial principles of ML is comprehending the data you are working with. These design principles are most important at the stage when the data are fed into the chosen algorithms (see, for example, \cite{Muller2016}). That is why we used a representative data set of the 315\,782 SDSS DR9 galaxies at $z<0.1$ with the absolute stellar magnitudes $-24^{m}<M_{r}<-13^{m}$, which we name as the target data set (see, in detail, Paper II \cite{Vavilova2022}). 

It was studied by us practically as ``galaxy by galaxy'' in previous works for various tasks (\cite{Melnyk2012, Dobrycheva2014, Dobrycheva2015, Dobrycheva2017, Dobrycheva2017a, Dobrycheva2018, Vasylenko2019, Khramtsov2019a, Vasylenko2020,  Vavilova2020a, Vavilova2021c}, including the ML photometry-based approach for binary galaxy morphological classification \cite{Vavilova2021a} and the catalog of their morphological types  \cite{Vavilova2021b} obtained with the Support Vector Machine and Random Forest methods. The paper II \cite{Vavilova2022} describes a general methodology for the CNN morphological classification as well as a morphological catalog of galaxies classified into five classes according to the Galaxy Zoo 2 labeling annotation is published through VizieR \cite{Vavilova2022b}.

\subsection{Training and inference data sets}

To provide the image-based approach for morphological classification of galaxies from the target data set, we used the GZ2 annotated data. To train the neural network, we should have a large number of labeled galaxies images. The target data set of the SDSS galaxies is tightly overlapped with the data from GZ2 \citep{Willett2013}. For this reason, we divided our target data set into two data sets. Hereafter, we determine the data set of 143\,410 galaxies, which do not match the GZ2 galaxies, as the ``inference'' data set. The data set of 172\,372 galaxies, which match the GZ2 galaxies, is the ``training'' data set. The sample from GZ2 contains all the well-resolved galaxies essentially in DR9 with $N$ = 11\,923 galaxies from the Stripe 82 (11.6 $\leq$ $m_{r}$ $\leq$ 17.7, $0.003 < z < 0.09$), where about of 6\,800 are at $0.07 < z < 0.09$. We considers galaxies only in normal-depth SDSS imaging and with DR9 spectroscopic redshifts.

We consider two types of morphological classification. The first type is the classification, which includes clearly separable five classes: completely rounded, rounded in-between, cigar-shaped, edge-on, and spiral galaxies. This classification is based on the combinations of precisely labeled GZ2 parameters and, obviously, includes only some part of the training data set. Unlike the first type, the second type of classification works with the 37 galaxy morphological features from the GZ2 and covers all galaxies presented in the training data set.

To form the first type of classification, we used specific criteria which allow us to separate different morphological classes of galaxies \citep{Willett2013}. These criteria were listed in the Paper ii. Besides, we removed seven galaxies which fit in two or more criteria listed in this Table. \footnote{The criteria with `\textsc{*\_count}' prefix indicate the number of votes; other criteria correspond to the debiased fraction of votes signed in the GZ2 catalogue as `\textsc{*\_debiased}'.} So, we exploited only those galaxies for training for which the most votes of GZ volunteers was collected. Such constraints are not all-inclusive. The more complete and severe criteria could be used to determine the morphological type of a galaxy as clearly as possible. However, as we discussed in the Paper II, the criteria in use are good enough to provide reliable image-based classification. 

To form the second type of classification (classification by the morphological features, down panel in Fig. \ref{fig:chart}), we used at least one of 37 features of galaxies from the training data set, which are described in the first column of Table \ref{tab:morphological_test1} and Table \ref{tab:morphological_test2}. Also, we removed three very sparse classes from the consideration ``bulge prominence dominant", ``odd feature lens or arc", ``bulge shape boxy") each containing of $<10$ galaxies. In total, we obtained the training data set of $160\,471$ galaxies (down panel, Fig. \ref{fig:chart}). To test the accuracy of the detailed morphological classification on the faint magnitude end, we also used 16\,626 galaxies from the DECaLS (see, subsection 3.5).  

\begin{figure}[ht]
\begin{subfigure}{.5\textwidth}
  \centering
  \includegraphics[width=1.\linewidth]{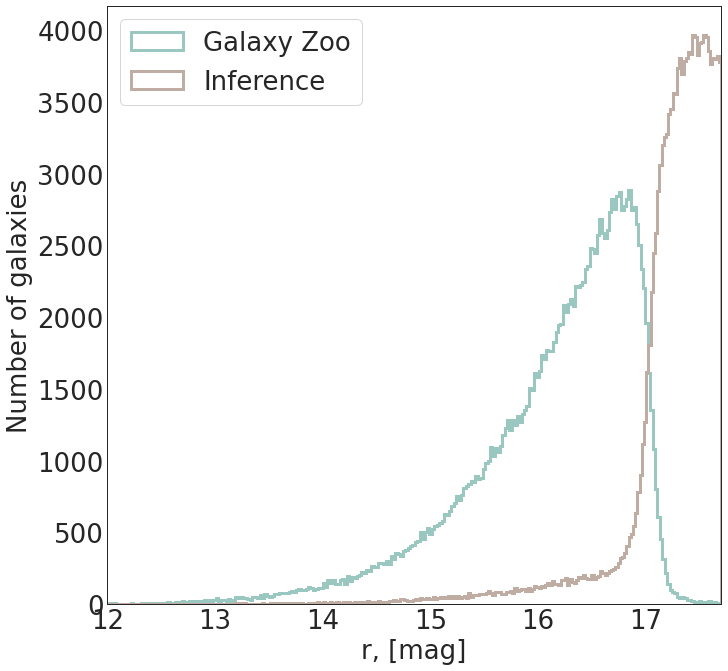}  
  \caption{The stellar magnitude.}
  \label{fig:rmag1}
\end{subfigure}
\begin{subfigure}{.5\textwidth}
  \centering
  \includegraphics[width=1.\linewidth]{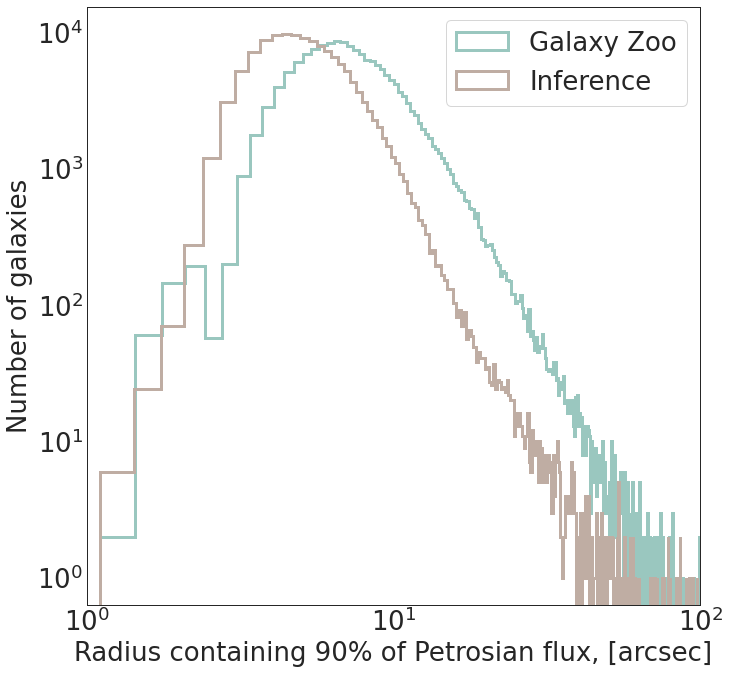}  
  \caption{The Petrosian radius (90\,\% of the flux).}
  \label{fig:rmag2}
\end{subfigure}
\caption{Histograms of the stellar magnitude and Petrosian radius (90\,\% of the flux) distributions in $r$-band for the training (green) and inference (brown) SDSS galaxy data sets at $z<0.1$.}
\label{fig:fig}
\end{figure}

There is a principal difference between galaxy images in our inference data set and training data set matching the GZ2 catalog. One can see in Fig.~\ref{fig:rmag1} and Fig.~\ref{fig:rmag2} that the inference data set is much shallower than the training one. This is occurred because the galaxies from the target data set of 315\,782 galaxies were pre-selected via $m_{r} < 17.7$ limitation by stellar magnitude in $r$-band. This limitation is related to the 90\,\% Petrosian flux parameter \cite{Blanton2001, Yasuda2001, Vavilova2021a}. So, the galaxies, which do not match the GZ2 catalog from the target data set, are, on average, fainter and smaller than galaxies from the training GZ2 data set. In total, 24\,547 galaxies from the inference data set have $m_{r} < 17.7$ (Fig.~\ref{fig:rmag1}). The CNN classifier knows nothing that it will works with the inference data set, where galaxies are fainter and smaller than in the training data set. So, it gives us an additional case to study performance of the image-based classification providing some additional steps.

Namely, to understand how crucial the shift between training and inference data sets is for the CNN classifier, we use additional test data set. It is based on the image morphological classification of 314\,000 galaxies from DECaLS and includes revealed fine morphological features, which are not seen with the SDSS images \cite{walmsley2021galaxy}. With this additional test data set, we identified  16\,626 galaxies in our inference data set, which further are used for the approach testing. We note that the morphological classification scheme for the DECaLS is slightly different from the one for the GZ2, namely it is biased towards increasing the detection of bars, measuring bulge size, and distinguishing types of merging galaxies. To align the GZ2 classification used in our study and the DECaLS morphological classification, we removed \textbf{15} classes from this data set because the DECaLS morphological classification does not contain some of the GZ2 classes (see, Table~\ref{tab:morphological_test1} and Table~\ref{tab:morphological_test2}). After this data preparation, we obtained 28 GZ2 features labels in our additional test data set. Hereafter in the paper, we name it as the ``deep" test data set.

\begin{figure*}
    \centering
    \includegraphics[scale=1.0]{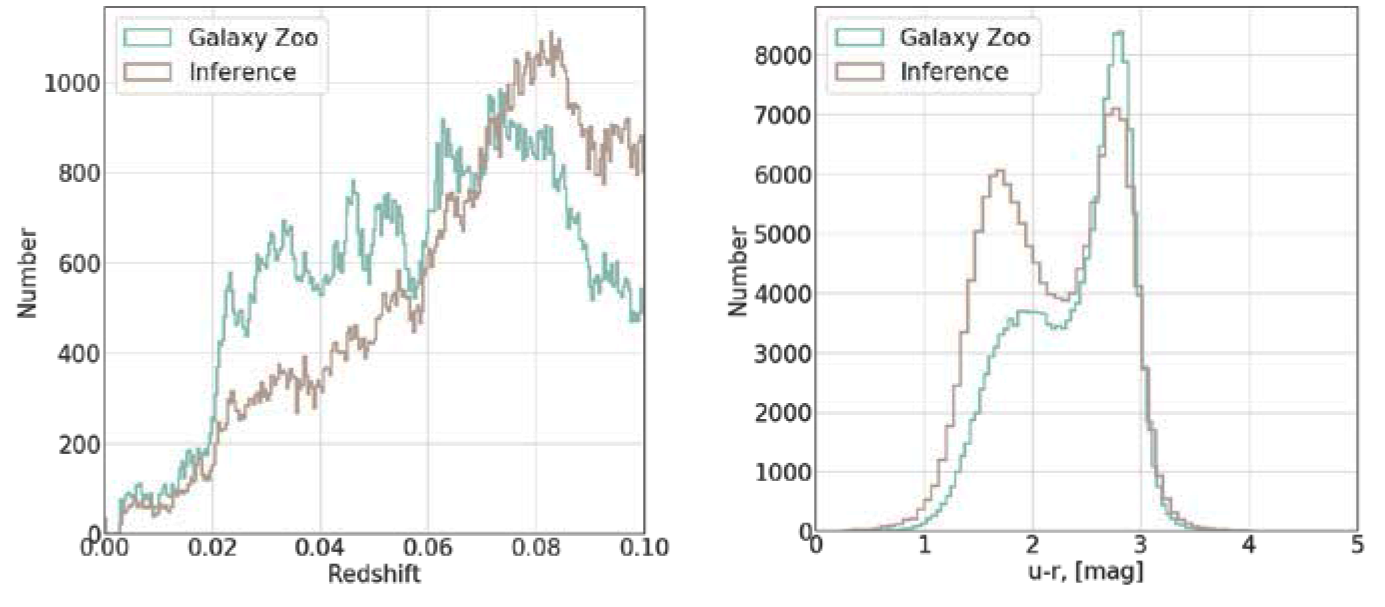}
    \caption{Histograms of the redshifts (left) and $u-r$ colour indices distributions for the training (green) and inference (brown) SDSS galaxy data sets at $z<0.1$.}
    \label{fig:redshift-color}
\end{figure*}

Other relevant observational parameters are better overlapped among two data sets, see, for example, Fig.\ref{fig:redshift-color} with distributions by redshift and $u-r$ color indices.

\subsection{Images of galaxies}

Images of the training and inference galaxies were requested from the SDSS cutout server\footnote{\url{http://skyserver.sdss.org/dr15/en/help/docs/api.aspx##cutout}}. We have retrieved 315,782 RGB images (in PNG format) composed of $gri$ bands according to \cite{Lupton2004} color scaling, each of 100$\times$100$\times$3 pixels\footnote{Corresponding to $39.6\times39.6$ arcsec in each channel of the RGB image.}. Unfortunately, some of the images were not retrieved \textbf{by technical reason (including dead pixels)}, slightly reducing the training and inference data sets to 172\,251, and 136\,342, respectively. 

We note that scientific image format (likely FITS) may be preferable in our task due to the higher amplitude ranges, in respect to 256 values per band in the simple PNG image. But such a flux sampling is more required for detailed image analyses, for example, gravitational lens modeling, while most of the deep-learning models are working on images with 8-bit amplitudes (see, for example, \cite{Wang2018}). Additionally, FITS files from the SDSS may be composed into 5-band images, expanding spectral information, while PNG files are restricted to have 3 bands only ($gri$ in our case). Investigation of this issue is out of scope for our paper, and we used standard approach of utilizing the SDSS image cutouts for galaxy morphological classification \cite{Willett2013}.

\subsection{Implementation}

All the deep-learning models were implemented using \textsc{PyTorch}\footnote{\url{https://github.com/pytorch/pytorch}} and \textsc{pytorch-image-models}\footnote{\url{https://github.com/rwightman/pytorch-image-models}} libraries. To train the models, we used GPU GeForce GTX 1080Ti.

\section{Methodology, CNN image-based galaxy classifier}\label{sec:approach}

We exploited CNNs to reveal the morphological classification of galaxies by their images. With this technique, we solve two different classification problems and handle a shift between training and inference data sets.

Usually, CNN consists of layers represented by a sequence of convolutional operations, activation functions, and pooling operations. The principal aim of the CNN is to find such convolutional kernels that are the result of applying the whole CNN to the image finalized in some target value. In our case, the morphological classes and features of galaxies are target values. The CNN architectures use the fully connected layers (instead of convolutional blocks) at the tail. This tail corresponds to the neural network classifier, which transforms the output of the convolutional part into the dense layer, the number of neurons, which is equal to the number of classes\footnote{A good practical overview can be accessed through \url{http://cs231n.stanford.edu/}.
We address readers also to  works \cite{Ren2014, Honghui2016, Flamary2016, Meyer2018}, where the feature extraction power of CNNs was illustrated in numerical experiments for improving the classification performance, including astronomical image reconstruction.} 

\begin{figure}[!ht]
    \centering
    \includegraphics[scale=0.3]{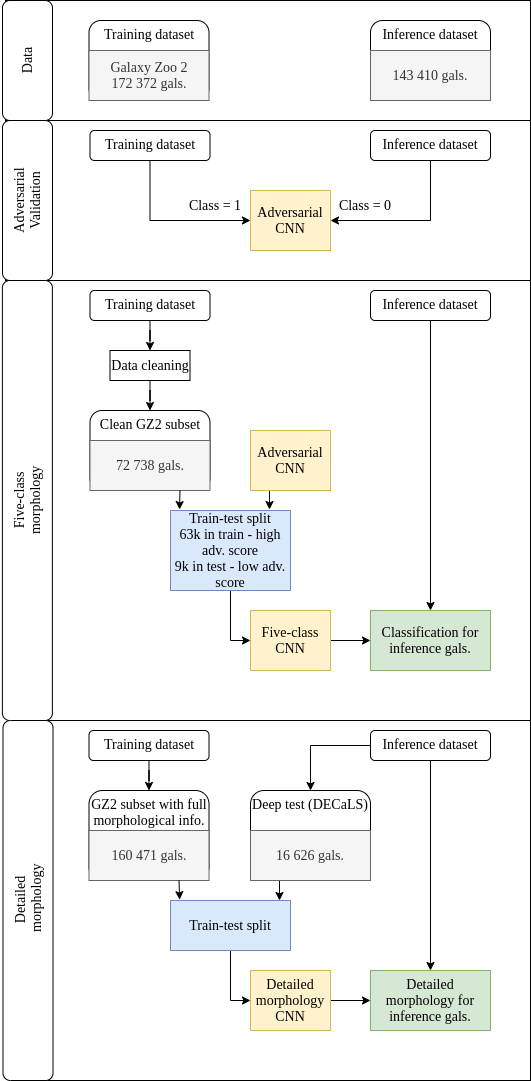}
    \caption{Scheme of the image-based approach for morphological classification of galaxies. Methodology consists of the data preparation, adversarial validation, five-class CNN morphological classification with intelligent train-test split via adversarial scores, and detailed  image feature  morphological classification.}
    \label{fig:chart}
\end{figure}

\subsection{General approach}\label{sec:app}

The scheme of our approach is shown in Fig.~\ref{fig:chart}. First, we divide the studied data set into the training and inference parts (Section~\ref{sec:data}). Since the inference data set is enormously different from the training one, we have to apply some necessary procedure with a final classification, namely the adversarial validation\footnote{This method is commonly used in data science competitions, see, e.g., \url{http://fastml.com/adversarial-validation-part-one/}; \cite{Pan2020}.}. It allowed us not only to probe the difference between the galaxy images in training and inference data sets (middle panel in Fig.~\ref{fig:chart}) but to derive the most suitable method of testing the CNN classifier, which will produce a representative estimation of quality on inference data set. This procedure is also significant in our approach for two reasons: the labeled galaxy data sets are biased in stellar magnitude distribution for the training data set (Fig. \ref{fig:rmag1}); such a difference could led to bias of the final prediction of galaxy classification in the inference data set. 

At the second stage of the pipeline, we use CNN to solve the five-class problem described in Section~\ref{sec:data}. We test our model with the data set defined by the adversarial validation. 

Finally, we train a second model to predict the detailed morphological features (likely bar, bulge, merging, ring, etc.), which is tested with the adversarial validation and deep test data sets. As the result of a pipeline, we obtain five morphological classes and 34 detailed morphological parameters for galaxies from the inference data set (down panel in Fig.~\ref{fig:chart}).

\subsection{Data preparation and augmentation}\label{sec:augment}

Stable CNN learning presumes the right scaling or normalization of the input data \citep{Bishop1995}. We scaled each image $I$ (pixels of which contain values between 0 and 255: $I_{i,j}\in\{0,255\}\:$) to the range $[-0.5,0.5]$ using the scaling equation as follows:

\begin{equation}
\centering
\tilde{I}_{i,j} = \frac{I_{i,j}-127.5}{255}
\end{equation}

Also, we defined many affine transformations for applying to images of galaxies during the CNN learning (so-called image augmentation). In our case, the augmentation helps to introduce the variative nature of galaxies to the CNN methods (because the standard CNNs are not fully invariant to any transformation of the images and have a strong ability to over-fitting). In most cases, this trick improves the generalization ability of CNN producing a less over-fitted model on the training data set (see, e.g. \cite{Dieleman2015}). 

As augmentations, we used random rotation (0$^\circ$, 90$^\circ$, 180$^\circ$, or 270$^\circ$), random zoom (varying at $100\div120$ pixels in each axis) with further random cropping of the $100\times100$ region, and random vertical/horizontal flipping of the images of galaxies. This process was applied randomly to each image of a galaxy so that each image of a certain galaxy was put in the CNN as a ``new'' one reducing the sensitivity of CNN to any galaxy orientation.

These augmentation steps were exploited during the adversarial validation with the CNN classification. We note in advance that after the adversarial validation was produced, we conducted additional data augmentation procedures that helped to learn the CNN classifier better (Section \ref{sec:adversarialvalidation}). 

\subsection{Adversarial validation}\label{sec:adversarialvalidation}

Having the training and inference data sets (Section~\ref{sec:data}), we can investigate how the images of galaxies ``vary'' between these data sets. We trained the CNN on all of these images, passing the class ``0" for inference data set and class ``1" for the training one (second panel, Fig.~\ref{fig:chart}).

In this case, the CNN classifier tried to distinguish the training images from images of galaxies from the inference data set, returning the ``adversarial score'' -- the probability of the galaxy being in the training data set. If such a classification accuracy is close to random guessing, one could assume the similarity of the training galaxy images with the inference ones. Moreover, vice versa, when the adversarial classification accuracy largely differs from random guessing (tends to the $100\%$), one has to investigate the difference between the training data set and the inference one to predict the classes of inference objects correctly.
Adversarial score is a measure of how an individual galaxy is similar to the training data set (larger scores correspond to larger similarities with galaxies from the training data set). The effect of dissimilarity is due to the different observed parameters of galaxies from the training and inference data sets. We used the full GZ2 data set as a training data set (comprising of 172\,372 galaxies) with adversarial class ``1''.

We employed \texttt{ResNet-101} \cite{he2015deep} as a model, where the convolutional part was completed by the two layers of neurons with 128 and 2~neurons in each layer correspondingly. After the first layer of neurons, we put on the \texttt{Leaky Rectified Linear Unit} activation function. The last layer that returns the probabilities of being in the training or inference dataset was supplemented by the \texttt{softmax} activation function. As an optimizer, we used \textsc{Adam} with initial learning rate $5\times 10^{-3}$; the optimizer minimized the \texttt{categorical\_crossentropy} loss function. In this way, we tried a single ResNet-101 model as a baseline approach and obtained a good accuracy for GZ2 vs inference classification. We did not vary models, because the aim is not to have a performance as higher as possible. The trained model is just a key-performance indicator for each galaxy, and its outputs were used as the proxy metric to understand similarity between the target (not GZ) data set and each galaxy or its augmented version. 

The whole input set consisted of $\sim$170\,000 galaxies from the GZ2 training data set and $\sim$136\,000 galaxies from the inference one. We have trained the model on 75\,\% of input data and validated it on the rest part of the galaxies. We applied standard data augmentation procedures to the training images described in Section~\ref{sec:augment}.
The model was learned during 12 epochs. If the overall classification accuracy of galaxy images from the validation data set did not increase during three epochs, we decreased the learning rate by a factor of $0.1$. Finally, we used the model that provided the best overall accuracy (91.28\,\% on the validation data and 91.67\,\% on the training one).

\begin{figure}[ht]
\begin{subfigure}{.5\textwidth}
  \centering
  \includegraphics[width=1.\linewidth]{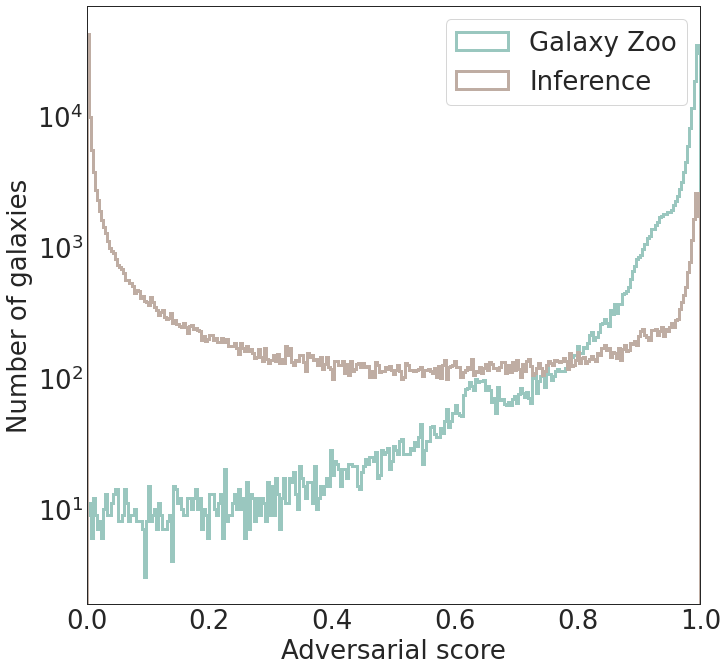}  
  \caption{The inference galaxy data set (brown) and for elliptical and spiral galaxies from the GZ2 training data set. Adversarial score is close to 1 if the galaxy is similar to the galaxy from GZ2 training data set.}
  \label{fig:adversarial_score}
\end{subfigure}
\begin{subfigure}{.5\textwidth}
  \centering
  \includegraphics[width=1.\linewidth]{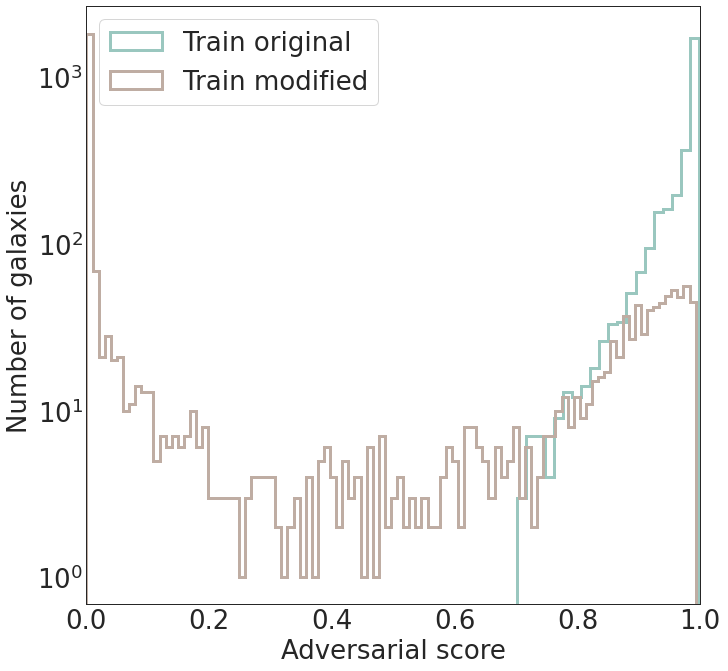}  
  \caption{A random subset of 3\,000 images from the training data set revealed from original SDSS images (green) and images with modified sizes and intensities of galaxies as $k=0.8$, $m=0.7$ from Eq.\ref{eq2} (brown).}
  \label{fig:adversarial_score_modified}
\end{subfigure}
\caption{Histograms of adversarial score distributions.}
\label{fig:fig}
\end{figure}

For our task, we obtained the accuracy of adversarial classification above $90\,\%$. So, the inference dataset contains galaxies with morphological properties which are not inherited to the training set. One can see in Fig.~\ref{fig:adversarial_score} that the adversarial score is relatively high for a few galaxies only from the inference data set. This agrees with our observation that inference galaxies are fainter (Fig.~\ref{fig:rmag1} and smaller  (Fig.~\ref{fig:rmag2}) than galaxies from the training data set.

We highlight that the resulting adversarial classification accuracy is not a result of over-fitting. Specifically, we randomly split the GZ2 training plus inference data sets into two parts. One of which was used to train the adversarial CNN and another to validate it. The CNN scored the same adversarial accuracy for these subsets (91 \%).

So, according to the adversarial result, we can conclude that our training data set contains galaxies, properties of which are not common with the inference one. This means that any validation of the morphological classifier has to be done with the galaxies from the training data set, which have a low adversarial score. 

This is a typical danger case of over-fitting, when a ML model is well performed on the training data set but is not able to generalize to new, previously-unseen data. This effect may be controlled through the train-test splitting. In such a way, a portion of the data (called the test data) is set aside for using only to assess the performance of the trained model and is not included into the training data set. To do so, we randomly choose 9\,000 galaxies with an adversarial score higher than $0.7$ from the training data set of 72\,738 galaxies (comprising five different morphological classes). We picked up the best threshold $0.7$ with a simple search taking into account the largest accuracy (see, Fig. \ref{fig:adversarial_score}); other thresholds result in lower separation quality. Within this train-test split, the test part of training galaxies (9\,000) was used to validate the morphological CNN classifier, and the rest part of the galaxies (63\,738) to train CNN classifier (middle panel, Fig. \ref{fig:chart}). It allows to understand the CNN ability to generalize on data it has never seen before, namely on the galaxies, which are similar to the inference dataset according to their adversarial score.

To train the CNN classifier for the prediction of the classes of fainter and smaller galaxies, we have added the following transformations of images to the defined data augmentation procedures (see subsection~\ref{sec:augment}):

\begin{equation} \label{eq2}
    \tilde{I}_{i,j} = \frac{k\times S(I_{i,j}, m)-127.5}{255}
\end{equation}

where $S(I_{i,j}, m)$ is a function changing the size of the image in $m$ times, and $k$ is an intensity-scaling coefficient. 

We implemented the size-changing function as simple zooming out of the image (into the new image with axes $(100\times m)\times(100\times m)$ pixels, where $0<m<1$), followed by mirror reflection of the image to fill up the missing $100\times(1-m)$ pixels along the borders. In turn, the intensity of pixels for each image was reduced by a factor of $0<k<1$. 

The augmentation procedures we implemented allow us to transform the image of the galaxy, simulating a decrease in magnitude and size as well as veiling it as the galaxy image from the inference dataset. For example, applying these transformations ($k=0.8$, $m=0.7$) to the 3\,000 random images from the training data set with adversarial score $>0.7$, we observed the shift of the adversarial score distribution towards zero value (see Fig.~\ref{fig:adversarial_score_modified}). The histogram of adversarial score distribution, especially for lower values, gives a direct confirmation in the support of such transformations. This trick with image transformations improves the accuracy of the result emulating the training galaxies to be more similar to the galaxies from the inference data set according to the adversarial scores. In this way, we do not investigate effects caused by different ``modalities” (training / inference); instead, we built a technique to prevent prediction biases. In other words, we solved the domain adaptation problem but with manually in-built heuristics (changing angular sizes and intensity of images of training galaxies).

\subsection{CNN five-class morphological classifier}

The next step of our pipeline was the morphological classification with CNN on training galaxy images. The principal difference between our approach and the existing ones (see, for example, recent works \cite{Walmsley2020, Vega2021, Bhambra2022, Gupta2022}) is the usage of (third panel, Fig. \ref{fig:chart})

1) the pre-defined training-test split through adversarial validation of the classification accuracy on the inference-like test set, and 

2) the specific data augmentation, which allowed us to decrease the difference in galaxy images related to the stellar magnitudes between the GZ2 and inference data sets. 

The procedure of training the CNN with the overall accuracy of 89.3\% on the test data set of 9\,000 galaxies is described in the Paper II. As for the data augmentation procedures, we used the standard techniques described in Section~\ref{sec:augment} and the intensity and size reduction of the galaxy images described in the previous subsection. For each galaxy from the training data set, we randomly sampled $k$ and $m$ from the uniform distribution within $(0.6,1.0)$ and $(0.5,1.0)$ ranges (Eq. \ref{eq2}), respectively. Data augmentation was applied to the training data set only. The confusion matrix of the distribution of predictions probabilities of all the classes is in Table 1 of the Paper II. One can see that an accuracy is not dominated by scores for morphological classes with higher numbers of galaxies.

Similar to the model for the adversarial validation, the tail of CNN models was completed by the two dense layers of neurons (with the number of neurons equal to 128 and 5, respectively) followed after the global max-pooling. The activation functions at the tail of the CNN model were the same as in adversarial validation. As an optimizer, we also used the \textsc{adam} with initial learning rate $1\times 10^{-4}$; the optimizer minimized the \texttt{categorical\_crossentropy} loss function. The CNN models were trained during 40 epochs. Moreover, during the learning, we decreased the learning rate value if the loss on the validation dataset was not decreased after four epochs. The eventual classification accuracy of the validation data set for all models is shown in Table~\ref{tab:morpho_architectures}. As the result of a comparison between these models, we decided to use \texttt{DenseNet-201} \cite{huang2018densely}, which shows the highest accuracy on the ``unseen'' validation (96.6\%) and test (89.3\%) data sets.

\begin{table}
    \centering
\begin{tabular}{l|c}
    Architecture & Accuracy \\
    \hline
    \texttt{ResNet-50$^a$} & 0.821 \\
    \texttt{ResNet-101$^a$} & 0.832 \\
    \texttt{ResNet-152$^a$} & 0.826 \\
    \texttt{InceptionV3$^b$} & 0.937 \\
    \texttt{InceptionResNetV2$^c$} & 0.962 \\
    \texttt{DenseNet-121$^d$} & 0.960 \\
    \texttt{DenseNet-169$^d$} & 0.959 \\
    \texttt{DenseNet-201$^d$} & 0.966 \\
    \texttt{NASNetLarge$^e$} & 0.929 \\
    \texttt{VGG16$^f$} & 0.909 \\
    \texttt{Xception$^g$} & 0.956 \\
\end{tabular}
\caption{Accuracy scores of backbone models for the five-classes of CNN morphological classification on the validation data. References: 
$^a$\cite{he2015deep}, 
$^b$\cite{szegedy2015rethinking}, $^c$\cite{szegedy2016inceptionv4}, $^d$\cite{huang2018densely}, 
$^e$\cite{Zoph2017}, 
$^f$\cite{simonyan2015deep}, 
$^g$\cite{chollet2017xception}
}
    \label{tab:morpho_architectures}
\end{table}

\subsection{Detailed galaxy morphology classification}

We used another CNN model to predict 34 detailed morphological parameters of galaxies from the inference data set. This model exploited \texttt{DenseNet-201} \cite{huang2018densely} as the backbone model with the included fully connected layers at the top (namely, global max-pooling, fully-connected layer with 512 neurons, and classification fully-connected layers with 34 outputs). We put \texttt{Rectified Linear Unit} activation after the first fully-connected layer, and \texttt{sigmoid} activation after the last classification of fully-connected layers. The model was trained with \textsc{adam} optimizer, which minimized the \texttt{binary\_crossentropy} loss function. We solved a multi-label classification problem: one object may have a few features. So we did not use the \texttt{softmax} activation function after the classification layer; instead, we treated each class separately and solved binary-classification for each label. This configuration looks suitable for solving \textit{multi}-label problem when we do not need to predict probability distribution over all classes to infer the single class for a single sample (forth panel, Fig. \ref{fig:chart}). 

We provide below in Table \ref{tab:morphological_test1} and Table \ref{tab:morphological_test2} two resulting accuracy scores measured with ROC AUC classification quality metric \cite{Bradley1997} to predict 34 morphological features of galaxies. The names of features are in the first column. The next columns correspond to the quality metrics (ROC AUC), ROC$^{\text{test}}$ for GZ2 test data set, ROC$^{\text{deep test}}$ for the DECaLS. We provided three scores for each data set: with and without adversarial augmentation, and the difference between both scores. The last three columns: threshold; number of galaxies matching this criterion from all the target data set and the inference data set, respectively. Empty cells correspond to the missed features. The sum numbers in columns 8 or 9 may not be equal to the total number of galaxies: one galaxy can have features in several classes, it is also possible that there are galaxies that do not fit any criterion.

These tables allow comparing this score for the model trained with adversarial augmentations (Section~\ref{sec:augment}) and for the model trained without these augmentations. Such a comparison should be useful to understand the degree of influence of image augmentations on the classification quality of the trained model. Scores are given for two test data sets: 1) for the data set of 9\,000 galaxies and 2) for the DECaLS galaxy data set. As one can see in these Tables, for the case of the GZ2 test data set, the scores, in general, are lower on tests for the model, trained with ``flux weakening'' and ``size reduction'' augmentations. This effect is explained by the similarity of the train and test data sets because due to the object selection in the GZ2 project, we are not able to sample a satisfactory amount of faint and small galaxies to test on. And thus, our adversarial augmentations shifts the training data set distribution with respect to the test data set.

At the same time, we note the improvement in the classification of the DECaLS galaxies. The scores overall are much lower than in the case of our GZ2 test data set. It may be explained by the revealing more fine structure of morphology with DECaLS: galaxies, which have some class in the GZ2, may be classified in another class with the DECaLS. But applying a model trained with adversarial augmentations leads to increasing the classification quality (except \textsc{star\_or\_artifact} class).

\begin{landscape}
\begin{table*}
    \centering
\begin{tabular}{l|c|c|c|c|c|c|c|c|r}
\hline\hline
Parameter & ROC$^{\text{test}}_{\text{NOAUG}}$ & ROC$^{\text{test}}_{\text{AUG}}$ & ROC$^{\text{test}}_{\text{diff}}$ & ROC$^{\text{deep test}}_{\text{NOAUG}}$ & \textbf{ROC}$^{\textbf{deep test}}_{\textbf{AUG}}$ & ROC$^{\text{deep test}}_{\text{diff}}$ &  Thre- & N, all  & N, inf. \\
 &  & & &  & &  &   shold & data & data\\
\hline    
\textsc{smooth} & 89.25\% & 88.59\% & -0.66\% & 86.06\% & 86.84\% & 0.78\% & 0.1 & 107\,657 & 51\,911\\
\textsc{features\_or\_disk} & 92.54\% & 91.88\% & -0.66\% & 85.63\% & 85.43\% & -0.20\% & 0.3 & 138\,207 & 58\,796  \\
\textsc{star\_or\_artifact} & 95.36\% & 97.63\% & 2.28\% & 57.70\% & 51.43\% & -6.27\% & 0.05 & 220 & 73   \\
\textsc{edgeon\_yes} & 98.81\% & 98.65\% & -0.16\% & 87.35\% & 88.26\% & 0.91\% &  0.05 & 34\,420 & 14\,489 \\
\textsc{edgeon\_no} & 97.21\% & 96.82\% & -0.39\% & 75.53\% & 76.41\% & 0.88\% & 0.25 & 72\,843 & 19\,088  \\
\textsc{bar} & 93.99\% & 92.41\% & -1.57\% & 57.54\% & 57.54\% & 0.00\% & 0.05 & 29\,892 & 6\,276 \\
\textsc{no\_bar} & 90.69\% & 89.80\% & -0.90\% & 68.82\% & 68.61\% & -0.21\% & 0.2 & 86\,836 & 27\,861  \\
\textsc{spiral} & 93.40\% & 92.88\% & -0.52\% & 78.97\% & 79.48\% & 0.51\% & 0.15 & 65\,709 & 17\,741\\
\textsc{no\_spiral} & 86.30\% & 84.78\% & -1.52\% & -- & -- & -- & 0.05 & 69\,303 & 20\,603  \\
\textsc{no\_bulge} & 98.36\% & 98.35\% & -0.01\% & 65.09\% & 69.03\% & 3.94\% & 0.05 & 6\,970 & 4\,046 \\
\textsc{bulge\_just\_noticeable} & 90.89\% & 89.75\% & -1.14\% & -- & -- & -- & 0.05 & 39\,627 & 14\,926 \\
\textsc{bulge\_obvious} & 90.55\% & 89.07\% & -1.49\% & 62.45\% & 64.31\% & 1.86\% & 0.05  & 27\,115  &  10\,018\\
\textsc{bulge\_dominant} & -- & -- & -- & -- & -- & -- & -- & -- & -- \\
\textsc{odd\_yes} & 94.78\% & 93.37\% & -1.41\% & -- & -- & -- & 0.05 & 41\,334 & 17\,601 \\
\textsc{odd\_no} & 84.62\% & 83.51\% & -1.11\% & -- & -- & -- & 0.45  & 170\,898  & 79\,134   \\
\textsc{completely\_round} & 96.17\% & 95.60\% & -0.58\% & 93.09\% & 93.51\% & 0.43\% &  0.15 & 75\,844 & 35\,669 \\
\textsc{rounded\_in\_between} & 92.31\% & 91.46\% & -0.85\% & 82.73\% & 82.84\% & 0.11\% &  0.2 & 125\,734 & 70\,389  \\
\textsc{cigar\_shaped} & 97.96\% & 97.73\% & -0.23\% & 97.24\% & 97.46\% & 0.22\% & 0.1 & 60\,395 & 30\,351  \\
\textsc{ring} & 96.97\% & 96.43\% & -0.54\% & -- & -- & -- &0.05 &  13\,882 &  1\,346 \\
\textsc{lens\_or\_arc} & -- & -- & -- & -- & -- & -- & -- & -- & -- \\
\end{tabular}
    \caption{Quality of inference morphological feature on the test data sets of galaxies (see, description in text of this paper). }
    
    \label{tab:morphological_test1}
\end{table*}
\end{landscape}

\begin{landscape}
\begin{table*}
    \centering
\begin{tabular}{l|c|c|c|c|c|c|c|c|r}
\hline\hline
Parameter & ROC$^{\text{test}}_{\text{NOAUG}}$ & ROC$^{\text{test}}_{\text{AUG}}$ & ROC$^{\text{test}}_{\text{diff}}$ & ROC$^{\text{deep test}}_{\text{NOAUG}}$ & \textbf{ROC}$^{\textbf{deep test}}_{\textbf{AUG}}$ & ROC$^{\text{deep test}}_{\text{diff}}$ &  Thre-  & N, all  & N, inf. \\
 &  & & &  & &  & shold   &data & data\\
\hline    
\textsc{disturbed} & 72.27\% & 68.55\% & -3.72\% & -- & -- & -- & 0.15  & 0  & 0 \\
\textsc{irregular} & 96.74\% & 96.94\% & 0.20\% & -- & -- & -- & 0.05 & 9\,432 & 6\,369  \\
\textsc{other} & 95.93\% & 89.20\% & -6.74\% & -- & -- & -- & 0.05 & 1\,442 & 624 \\
\textsc{merger} & 91.79\% & 88.89\% & -2.90\% & -- & -- & -- & 0.05 & 2\,575 & 738  \\
\textsc{dust\_lane} & 99.39\% & 99.40\% & 0.02\% & -- & -- & -- & 0.05 & 588 & 67  \\
\textsc{bulge\_shape\_rounded} & 96.73\% & 96.27\% & -0.47\% & 67.18\% & 67.26\% & 0.08\% & 0.05 & 32\,280 & 12\,835  \\
\textsc{bulge\_shape\_boxy} & -- & -- & -- & -- & -- & -- & -- & -- & --   \\
\textsc{bulge\_shape\_no\_bulge} & 98.65\% & 98.52\% & -0.13\% & 71.61\% & 71.46\% & -0.16\% &  0.05 & 19\,570 & 10\,867\\
\textsc{arms\_winding\_tight} & 89.45\% & 88.60\% & -0.85\% & 72.25\% & 72.29\% & 0.04\% &0.05 & 22\,180 & 5414  \\
\textsc{arms\_winding\_medium} & 75.33\% & 77.59\% & 2.26\% & 69.91\% & 71.57\% & 1.66\% & 0.05 & 304 & 86 \\
\textsc{arms\_winding\_loose} & 94.95\% & 94.41\% & -0.54\% & 69.03\% & 69.95\% & 0.92\% & 0.05 & 8\,411 & 3\,269  \\
\textsc{arms\_number\_1} & 85.56\% & 83.30\% & -2.26\% & 60.22\% & 61.83\% & 1.61\% & 0.05 & 445 & 188\\
\textsc{arms\_number\_2} & 90.55\% & 89.99\% & -0.56\% & 76.33\% & 76.62\% & 0.30\% & 0.05 & 69\,229 & 22\,061  \\
\textsc{arms\_number\_3} & 93.54\% & 93.47\% & -0.07\% & 70.14\% & 68.55\% & -1.58\% &  0.05 & 889 & 78  \\
\textsc{arms\_number\_4} & 93.84\% & 85.45\% & -8.39\% & 54.95\% & 56.96\% & 2.01\% & 0.05 & 82 & 3 \\
\textsc{arms\_number\_more\_than\_4} & 97.79\% & 97.51\% & -0.27\% & -- & -- & -- & 0.05 & 55 & 4  \\
\textsc{arms\_number\_cant\_tell} & 86.13\% & 86.07\% & -0.06\% & -- & -- & -- &0.05 &  7\,683 &  1\,329 \\
\end{tabular}
    \caption{(continue) Quality of inference morphological feature on the test data sets of galaxies. }
    \label{tab:morphological_test2}
\end{table*}
\end{landscape}

\section{General results and Discussion}

There are many classifiers for sorting galaxies by morphological type and features, but each has its own drawbacks. For example, spectroscopy classification requires different methods to define simultaneously similar spectra for quiescent/starburst and star-forming galaxies \cite{Rahmani2018, Curti2022} or emission-line galaxies \cite{Shi2015}. As well, a photometry-based approach gives an error when trying to classify red spirals and blue ellipticals (\cite{Tempel2011, Tojeiro2013, Vavilova2015, Vasylenko2019, Guo2020, Vavilova2020a}), i.e., galaxies with a high content of old stars or interacting galaxies which affect the photometric characteristics of each other (\cite{Mezcua2014, Simmons2017, Bottrell2019, Pearson2019}). Analyzing our obtained results and data products let us to discuss several issues related to the CNN image-based galaxy classification.

\subsection{Accuracy}

We applied CNN classifier to the studied low-redshifts SDSS galaxies and seized two sets of parameters: predictions of beings in one of five classes and to have one of 34 detailed morphological features using the GZ2 labeling. We remind that the five GZ morphological classes are relevant to the certain galaxy morphological types, e.g., T-types by de Vaucouleurs. Also, the human bias, which is caused by the GZ volunteers' answers in decision tree, affects the classification accuracy. It is discussed by many authors in different aspects (see, if interesting, ``Astronomy Blog. Galaxy Zoo and human bias"\footnote{\url{https://www.strudel.org.uk/blog/astro/000758.shtml}}.  We refer to the paper by Cabrera et al. \cite{Cabrera2018}, where the metric for human labeling measuring in case of low-redshift spiral/elliptical galaxies is proposed in frame of label's comparison between experts, GZ volunteers, and ML models. Hart et al. \cite{Hart2016} developed a reliable method for defining spiral galaxies, which eliminates the redshift-dependent bias in the GZ2 volunteer’s answers. It was taken into account ``by modeling the vote fraction distributions as a function of redshift, and correcting the higher redshift vote distributions to be as similar as possible to equivalent vote distributions at low redshift." 

We exploited the GZ2 annotated data as by Willett et al. \cite{Willett2013}, which can possess a worse bias for, as example, the late type galaxies (spiral) as compare with the data by Hart et al. \cite{Hart2016}, Of course, the exploiting more and more unbiased data for training should improve the accuracy of CNN classifier, see, for example, Tarsitano et al. \cite{Tarsitano2022}, where this debiasing technique is applied for ``disk and smooth" galaxies. Nevertheless of using the data \cite{Willett2013}, in general, our method is on par with most contemporary level of morphological classification performance attaining the accuracy of 83.3-99.4\% of depending on the image feature parameter within the GZ2 type questions (Table~\ref{tab:morphological_test1} and Table \ref{tab:morphological_test2}). Such an overall value of the accuracy is in a good agreement with the one obtained in work by Walmsley et al. \cite{Walmsley2020}, who used Bayesian CNN to study Galaxy Zoo volunteer responses and achieved coverage errors of 11.8 percent within a vote fraction deviation of 0.2. 

If consider the attained accuracy for certain morphological types of galaxies, we note the work by Gauthier et al. \cite{Gauthier2016}, who applied both supervised and unsupervised methods to study the Galaxy Zoo data set of 61\,578 pre-classified spiral, elliptical, round, and disk galaxies. They attained 94\,\% accuracy for galaxies to be associated with each of these four classes and noted the correlation of variation of galaxy images with brightness and eccentricity. Among other relevant works, we note by Barchi et al. \cite{Barchi2020} who used DL and traditional ML techniques for binary distinguishing of elliptical/spiral galaxies and created a morphological catalogue of 670\,560 galaxies at $z<0.1$, where the input data were taken from the SDSS DR7 (Petrosian magnitude in $r$-band $<17.78$). They developed a non-parametric galaxy morphology system (CyMorph). The Decision Tree, Support Vector Machine, and Multilayer Perceptron produced 98\,\% of overall accuracy. The CNN method (GoogLeNet Inception) with the imbalanced data sets and twenty-two-layer network resulted in 98.7\,\% overall accuracy for this binary morphological classification. Mitta et al. \cite{Mittal2020} introduced the data augmentation-based MOrphological Classifier Galaxy using CNN (daMCOGCNN) and obtained a testing accuracy of 98\,\% on the data sets of 4\,614 images from the SDSS, Galaxy Zoo challenge, and Hubble Image Gallery. 

\subsection{Train-test split. Transformation of images by intensity and size. Adversarial validation}

We revealed that adversarial validation is very helpful when the labeled data sets are biased in magnitude distribution for the training data set, and such a difference could bias the final prediction of the classifier on the inference data. So, we apply the adversarial validation method to analyze the homogeneity of the two data sets (inference and training). As a result, the galaxies were selected from the training data set that most closely coincided with the inference data set, and the images were normalized to be similar.

The principal difference of our approach is the pre-defined training-test split through adversarial validation of the classification accuracy on the inference-like test data set (Fig. \ref{fig:chart}). The deal with testing classification quality on different distributions (e.g., between training and target datasets) has a few implementations for galaxy morphology classifications (\cite{Sreejith2018, Ghosh2020, Walmsley2022, Gupta2022}). Below we note several of them.

Gauci et al. (\cite{Gauci2010}) used decision tree algorithms trained on $gri$ photometric information (colour indexes, shape parameters) to distinguish between spiral and elliptical galaxies or star/ unknown galactic objects from SDSS DR7 following the GZ annotated data.  They revealed that the incorrectly classified spiral and elliptical samples are very faint in magnitude. Our approach with adversarial augmentation and reveling differences between training and inference datasets allows to avoid this problem.

The transfer learning approach to fine-tune the CNN on dataset, different from the training one, has been recently acted by Ghosh et al. \cite{Ghosh2020} in their CNN classifier for bulge- and disk-dominated galaxies of the SDSS and Cosmic Assembly Near-Infrared Deep Extragalactic Legacy Survey (CANDELS). Inclusion of this procedure allowed them to overcome the problem of non-accurate predictions on the unseen datasets, with fine-tuning the network on the target dataset. Dominguez-Sanchez et al. created a morphological catalog for $\sim$670\,000 SDSS-galaxies in two options (T-type, related to the Hubble sequence, and GZ2 types). They obtained the highest accuracy ($>97$\%), when applied the same parameters to a test data set as the one used for training data set \cite{Dominguez2018}. 

But the labeled data from the target distribution is an essential condition to conduct the transfer learning. We handled this limitation simply by the imposing the required transformations into the training dataset, preventing the need of labels for target galaxies. 

Lin et al. \cite{YaoYuLin2021} used the Vision Transformer model, which operates better at classifying smaller-sized and fainter galaxies (in comparison to the CNN). This improvement is caused, probably, by the architecture change from the CNN to the attention-based model -- because transformers usually work better with a training dataset increasing, and, at the same time, these challenging types of galaxies were dominated in their training dataset. Lin et al. \cite{YaoYuLin2021} applied thresholds on a series of voting GZ2 questions \cite{Willett2013}, but considered eight classes: round elliptical, in-between elliptical, cigar-shaped elliptical, edge-on, barred spiral, unbarred spiral, irregular, and merger on the data set of 155\,951 galaxy images, obtaining the accuracy (with equal class weights) from 68.7 \% to 90.7 \% in dependence on the class excepting irregular (41.3 \%) and mergers (53.1 \%).

Dieleman et al. \cite{Dieleman2015} used similar data augmentation when provided the GZ decision tree model to predict probabilities for each of 34 answers of the GZ volunteers for the evaluation set of 79\,975 SDSS galaxy images. They selected the subset of images for which at least 50\% of volunteers answered the question. Exploiting translational and rotational invariation of galaxy images via data augmentation and keeping the centre of the galaxy as the most informative part, they also used random rescaling, flipping, and brightness adjustment. For images with high agreement among the GZ participants, their model provides an accuracy of more than 99\% for most questions.

The aforementioned results show a success of a standard data augmentation technique, while sophisticated augmentations -- to adapt the training set to the inference one -- are also effective, as we demonstrate in this paper.

\subsection{CNN classification by five morphological classes of galaxies}

\begin{figure*}[!ht]
     \centering
     \includegraphics[scale=0.35]{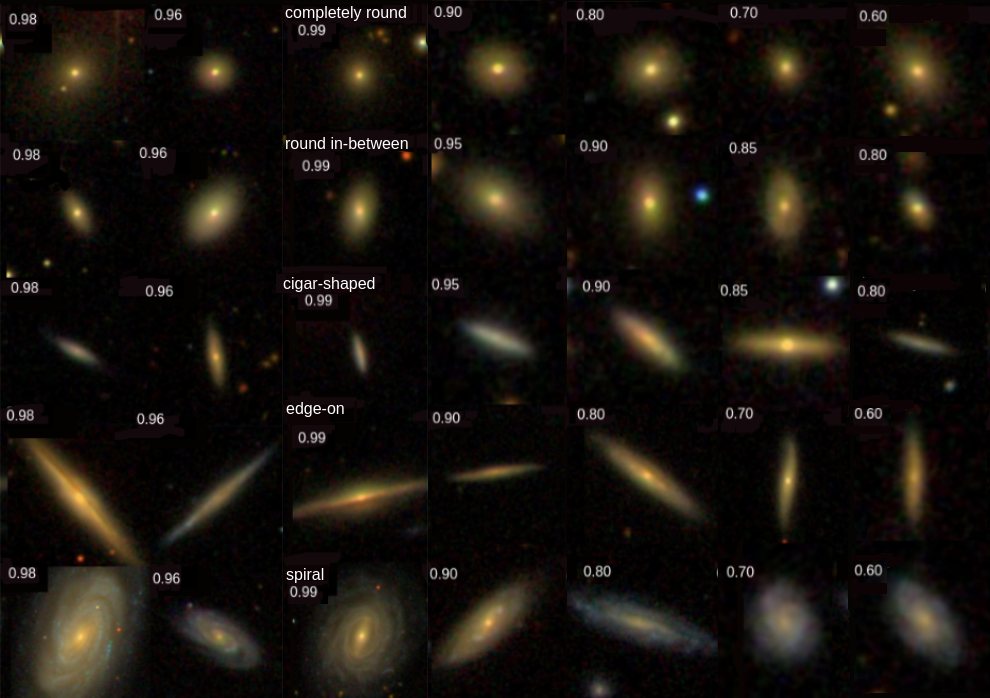}
     \caption{A set of the inference galaxies (3-7 columns) with their two nearest neighbours from the GZ2 training data set (1-2 columns). Each row represents the morphological class, which is intrinsic to the galaxy from the training data set. A value of the corresponding probability of being this galaxy in a given class is pointed in the left upper corner of each image. }
     \label{fig:composite}
 \end{figure*}

Assuming that a galaxy is in a certain class if the probability is the highest one, we have found that the inference data set comprises 27\,378 completely round (with probability of 83 \%), 59\,194 round in-between (93 \%), 18\,862 cigar-shaped (75 \%), 7\,831 edge-on (93 \%), and 23\,119 spiral (96 \%) galaxies (see, examples, in Fig. \ref{fig:composite}, similarity search). 
The Catalogue of 315\,776 SDSS DR9 galaxies at $z<0.1$ with image-based morphological classification by five classes is available through the UkrVO web-site \url{http://ukr-vo.org/starcats/galaxies/gal_SDSSDR9_z_to_0.1_morph_5_classes.csv} and VizieR \cite{Vavilova2022b} to be supplemented with the Paper II \cite{Vavilova2022}. It contains the CNN morphological classification of 72\,738 galaxies from the training GZ2 data set, 143\,410 galaxies from the inference data set (the faintest galaxies of the studied sample), 99\,528 galaxies from the GZ2 sample that did not pass the selection according to the criteria of the most votes of GZ2 volunteers and for which their morphological class was reassigned by the CNN classifier.

In this way, we have shown for the first time that implication of the CNN model with adversarial validation and size-changing function simulating a decrease in magnitude and size (data augmentation) significantly improves the classification of smaller and fainter SDSS galaxies with $m_{r} <17.7$ in $r$-band (Fig.~\ref{fig:adversarial_score}. One can see in Fig. \ref{fig:rmag1} that the fainter end of distribution of the target data set by magnitude is occupied by galaxies from the inference data set only. As well, we demonstrated another way to improve the human bias for those galaxy images that had a poor vote classification in the GZ project. Such an approach, likely auto-immunization, when the CNN classifier trained on very good images is able to retrain bad images from the same homogeneous sample, can be considered co-planar to other methods of combating the human bias, e.g. method proposed by Hart et al. \cite{Hart2016}.

It is relevant to compare our results with work by Zhu et al. \cite{Zhu2019}, in which ResNet model was exploited to classify galaxies into five classes annotated by GZ2 and compared CNN classifier with Dieleman et al. (\cite{Dieleman2015}), AlexNet, VGG, and Inception networks. The samples were pre-selected in a specific morphology category with their appropriate thresholds \cite{Willett2013} in dependence on the number of volunteers’ votes. These authors attained overall classification accuracy of 95.21 \% and the accuracy of each class type as 96.68 \% for completely round, 94.42 \% for round in-between, 58.62 \% for cigar-shaped, 94.36 \% for edge-on, and 97.70 \% for spiral. We had a comparable classification performance with a worse output for completely round and a better output for cigar-shaped classes. Gupta et al. \cite{Gupta2022} provided a classification of GZ2 galaxy images on five morphological classes as in our work. They trained Neural ordinary differential equations with Adaptive Checkpoint Adjoint and compared them against the ResNet CNN model: an accuracy of 91-95\% depending on the image class is in agreement with our results.

\begin{figure}[!ht]
    \centering
    \includegraphics[scale=0.25]{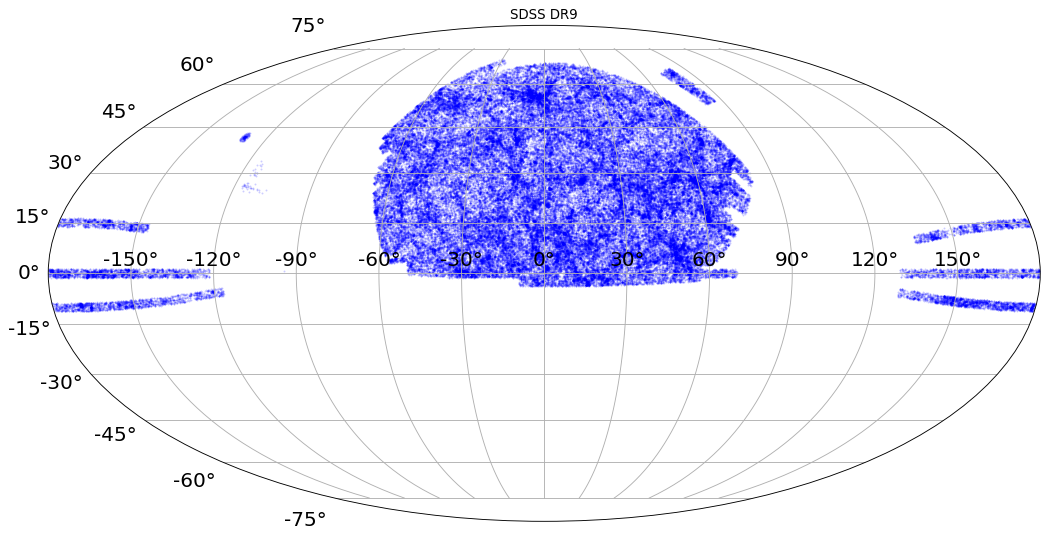}
    \caption{Distribution of galaxies classified by CNN as belonging to the round in-between morphological class in the sky.}
    \label{fig:radec-distr}
\end{figure}

\begin{figure}[!ht]
    \centering
    \includegraphics[scale=0.6]{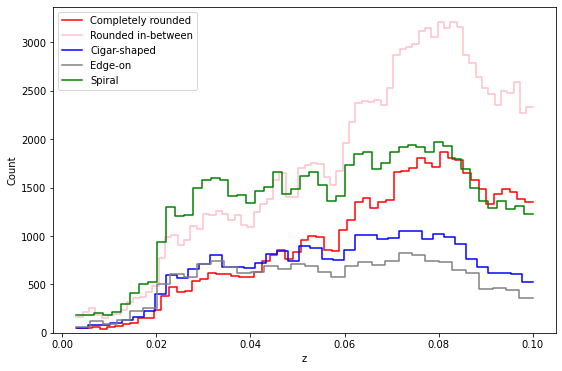}
    \caption{Distribution of galaxies classified by CNN into five morphological classes by redshift.}
    \label{fig:redshift-distr}
\end{figure}

Yet one point of the discussion is related to the distribution of galaxies on the sky and by redshift. For example, Dhar and Shamir \cite{Dhar2022} demonstrated that the training of a deep CNN is sensitive to the context of the training data such as the location of the objects on the sky. They found statistically significant bias in the form of cosmological-scale anisotropy in the distribution of elliptical and spiral galaxies, which affect deep CNN model. They experimented with Pan-STARRS and SDSS data and noted that such unbalancing is linked to the training and test samples of galaxies, which were imaged in different parts of the sky. We analyzed distribution of galaxies in our catalogues and have not found that galaxies of a certain morphological class (or morphological feature) have a preferential distribution in their location in the sky (see, as example, Fig. \ref{fig:radec-distr} for the most numerous round-in between class and Fig. \ref{fig:redshift-color} (left) for the training and inference data sets. There is no differences between classes in distribution by redshift (Fig. \ref{fig:redshift-distr}).

\begin{table}[]
\begin{tabular}{|c|c|c|c|c|c|c|c|}
\hline
\multirow{3}{*}{\textbf{\begin{tabular}[c]{@{}c@{}}Photo\\metry-\\ based\end{tabular}}}  & \multirow{3}{*}{\textbf{N gal}} & \multirow{3}{*}{\textbf{Type}} & \multicolumn{5}{c|}{\textbf{Image-based, \textbf{CNN}, classes}} \\ \cline{4-8} 
  &   &  & \multicolumn{1}{c|}{\multirow{2}{*}{\textbf{\begin{tabular}[c]{@{}c@{}}complete\\ly round \end{tabular}}}} & \multicolumn{1}{c|}{\multirow{2}{*}{\textbf{\begin{tabular}[c]{@{}c@{}}round in\\between\end{tabular}}}} & \multicolumn{1}{c|}{\multirow{2}{*}{\textbf{\begin{tabular}[c]{@{}c@{}}cigar- \\ shaped\end{tabular}}}} & \multicolumn{1}{c|}{\multirow{2}{*}{\textbf{\begin{tabular}[c]{@{}c@{}}edge-\\ on\end{tabular}}}} & \multirow{2}{*}{\textbf{spiral}} \\
   &     &    & \multicolumn{1}{c|}{}      & \multicolumn{1}{c|}{}     & \multicolumn{1}{c|}{}   & \multicolumn{1}{c|}{}    &    \\ \hline
\multirow{3}{*}{\begin{tabular}[c]{@{}c@{}}{\textbf{MPD}},\\ N=308\,466\end{tabular}} & 138\,947      & E    & 35\,389   & 65\,839   & 14\,360     & 12\,067   & 11\,292   \\ \cline{2-8} 
                                                                         & 110\,454      & Sp   & 13\,645   & 41\,047   & 12\,803     & 6\,065    & 36\,894   \\ \cline{2-8} 
                                                                         & 59\,065       & Irr  & 7\,627    & 20\,658    & 4\,224     & 2\,108   & 24\,448        \\ \hline
\multirow{2}{*}{\begin{tabular}[c]{@{}c@{}}{\textbf{RF}},\\ N=308\,466\end{tabular}}  & 131\,663      & Early    & 36\,424 & 66\,043    & 12\,268    & 8\,549    & 8\,379    \\ \cline{2-8} 
                                                                         & 176\,803      & Late & 20\,237 & 61\,501       & 19\,119   & 11\,691  & 64\,255      \\ \hline
\multirow{2}{*}{\begin{tabular}[c]{@{}c@{}}{\textbf{SVM}},\\ N=308\,466\end{tabular}} & 131\,099   & Early    & 36\,135          & 65\,646    & 12\,477      & 8\,790   & 8\,051    \\ \cline{2-8} 
                                                                         & 177\,367   & Late   & 20\,526  & 61\,898  & 18\,910  & 11\,450    & 64\,583    \\ \hline
\end{tabular}
 \caption{Comparison of classifications of the studied SDSS DR9 galaxies by CNN model into five morphological classes and by three photometry-based methods: multi-parametric diagram (MPD) into elliptical, spiral, irregular galaxies \cite{Melnyk2012, Dobrycheva2015}; machine learning with Random Forest (RF) and Support Vector machine (SVM) into early and late morphological types \cite{Vavilova2022, Vavilova2022b}. The number of only those galaxies, N gal, that have the best threshold probability of belonging to one or another morphological class is pointed out.}
    \label{tab:morphological_test4}
\end{table}

To compare photometry-based and image-based approaches to the same data set of low-redshift galaxies, we collected the classification output of four methods in Table~\ref{tab:morphological_test4}. There are results of classifications by the CNN model into five morphological classes; photometry multi-parametric diagram (MPD) into elliptical, spiral, irregular galaxies; machine learning with Random Forest (RF) and Support Vector Machine (SVM) into early and late morphological types. We inserted the number of only those galaxies that have the maximum probability of belonging to one or another morphological class \cite{Vavilova2022b}. One can see that three photometry-based methods have comparable overall accuracy with intrinsic error less than 0.3\% between RF and SVM \cite{Vavilova2021a} as well as less than 4\%
between MPD (here late type is Sp+Irr) and machine learning methods. The latter error is explained mostly by affect of blue elliptical and red spiral galaxies \cite{Melnyk2012}. There is a general agreement between the early type of galaxies classified by photometry methods and ``round-in-berween + completely round" types of galaxies as well as between late-type galaxies and ``spiral + round in-between". 

We matched the galaxies of late morphological types classified by Support Vector Machine (SVM) and Random Forest (RF) \citep{Vavilova2021b} and the galaxies classified in this work by CNN as edge-on and spiral as the most relevant morphological types. Namely, we selected $\sim$50\,000 galaxies with CNN probability to be spiral from 0.77 to 0.99 (Table~\ref{tab:morphological_test1} and Table~\ref{tab:morphological_test2}). Their labeling obtained by SVM and RF methods says that 10.5\,\%, and 8.8\,\% among them, respectively, are of early morphological type (elliptical). We inspected these misclassified galaxies and found that they are mostly large nearest spiral galaxies with a massive red center region. We also selected $\sim$12\,000 edge-on galaxies with the same CNN probability: also having a redder colour and larger redshifts. 

The comparison in Table \ref{tab:morphological_test4} shows significant segregation of galaxies classified by five GZ2 morphological classes between the adopted morphological types. This complicates the work of the CNN classifier to reveal the real morphology of galaxies. The statistical comparison of these results with the results of the CNN detailed morphology of the same five classes (Table \ref{tab:morphological_test1}) is impossible because a feature-classified galaxy can have multiple features, while a class-classified galaxy belongs to only one class.

In our opinion, it is more efficient to use the existing catalogues of galaxies (for example, elliptical, spiral, irregular, flat, gravitational lenses, mergers, etc.) as training ones to determine the morphological types of galaxies. Binning these catalogues by redshift we can sequentially create new morphological catalogues at higher redshifts, and after a thorough check, to use newly catalogues as training, etc. The emergence of new data on galaxy images for more in-depth samples by future observatories will provide such an algorithm by the data for CNN models. Meanwhile, it is useful to use both photometry- and image-based methods. Our approach to transfer the annotated classification of galaxies to fainter and smaller galaxies using adversarial validation with train-test splitting and image sizing is in favor of the correct applicability of CNN classifier and the efficiency of the algorithm.

\subsection{CNN classification by the detailed galaxy morphological features}

\begin{figure*}[!ht]
     \centering
     \includegraphics[scale=0.35]{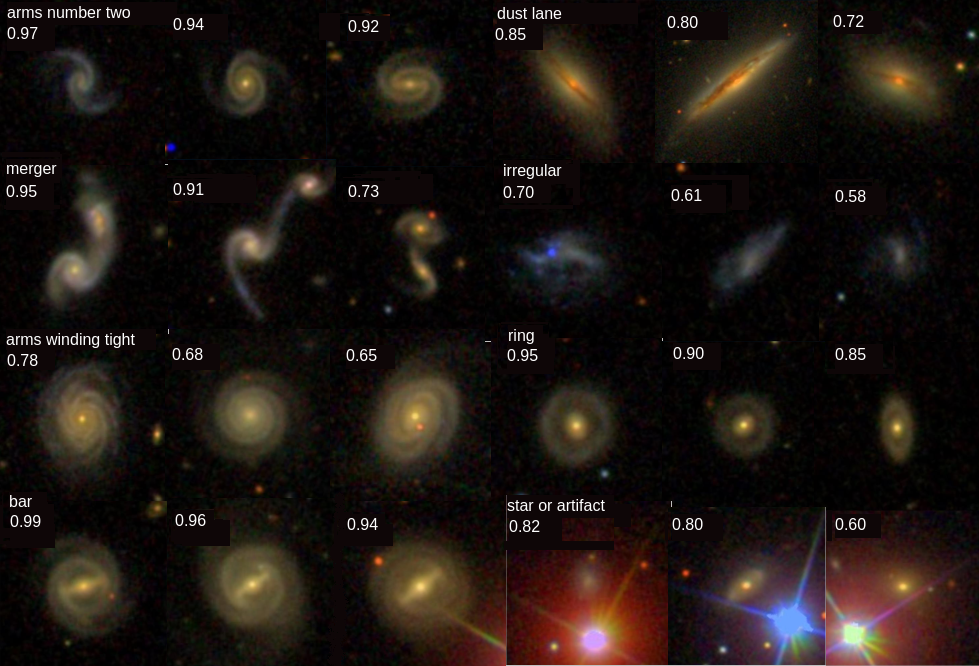}
     \caption{The examples of galaxies with some morphological features (bar, ring, irregular, merger, dust lane, arms winding tight, arms number 2, star or artifact) from the inference SDSS data set with their two nearest neighbors from the GZ2 training data set.}
     \label{fig:features}
 \end{figure*}

Quality of inference morphological feature on the test data sets of galaxies is summarized in Table~\ref{tab:morphological_test1} and Table \ref{tab:morphological_test2}. Our CNN model for the classification of galaxies by their detailed structural morphology gives the accuracy in the range of \textbf{83.3–99.4\%} depending on 32 features (exception is for ``disturbed" (68.55\%) and ``arms winding medium" (77.39\,\%), the number of galaxies with the given feature in the inference data set, the galaxy image quality (Table~\ref{tab:morphological_test1} and Table \ref{tab:morphological_test2}). To reach it, we calculated the number of galaxies that passed the \textbf{selected threshold} for the acceptance of detailed morphological features. The examples of classification on inference galaxy data set are given in Fig.~\ref{fig:features}. As a result, for the first time, we assigned the detailed morphological classification for more than 140\,000 low-redshift galaxies with $m_{r}<17.7$ from the SDSS DR9, which has the highest adversarial score by CNN classifier.

Using the adversarial validation technique, we managed the optimal train-test split of galaxies from the training data set to verify our CNN model based on the DenseNet-201 realistically. We have also found optimal galaxy image transformations, which help increase the classifier's generalization ability as is tested with a specifically created test data set. We can compare our results with work by Dieleman et al. \cite{Dieleman2015}. Namely, a level of agreement and model confidence presented in Fig. 9 of their paper demonstrates that classification overall accuracy for the analysed examples is in the range of 82.52-96.04\,\% in dependence on the galaxy feature (exception is for ``no of arms", ``arm tightness", ``odd", and ``bulge", where accuracy is less 80\,\%). Exploiting similar augmentation procedures for the SDSS galaxy images, our approach was slightly different: in the choice of image data as the PNG files restricted for three $gri$ bands as well as we conducted a multi-label task for detailed morphological classification, when the galaxy can be attributed with several features (for, example, labeling as ``spiral", the galaxy can be also with ``bar", ``bulge" or ``ring" and be characterized by certain number of ``arms").

A good train-test sampling mobility for CNN classifier is resulted in the catalogues of low-redshift galaxies with morphological features, which are supplemented to this paper. The highest score (97-99\,\%) was attained for such features as ring, irregular shape, bulge, star or artifact, edge-on, and dust lane.

So, we can underline that train/test split has very important consequences because with its use the CNN’s applicability to the future LSST, WFIRST, Euclid big data surveys will not depend on the need for a large training set of real data. 

In general, this allowing to make a quick selection of galaxies with certain features for their subsequent analysis (see Table~\ref{tab:morphological_test1} and Table \ref{tab:morphological_test2}). Using the SDSS Navigate, we performed a preliminary visual inspection of samples of galaxies with such features as ``dust lane, irregular, edge-on yes, ring, bar, merger, star or artifact" in order to reveal CNN efficiency to classify images from an astronomical point of view. 

\begin{figure*}[!ht]
     \centering
     \includegraphics[scale=0.25]{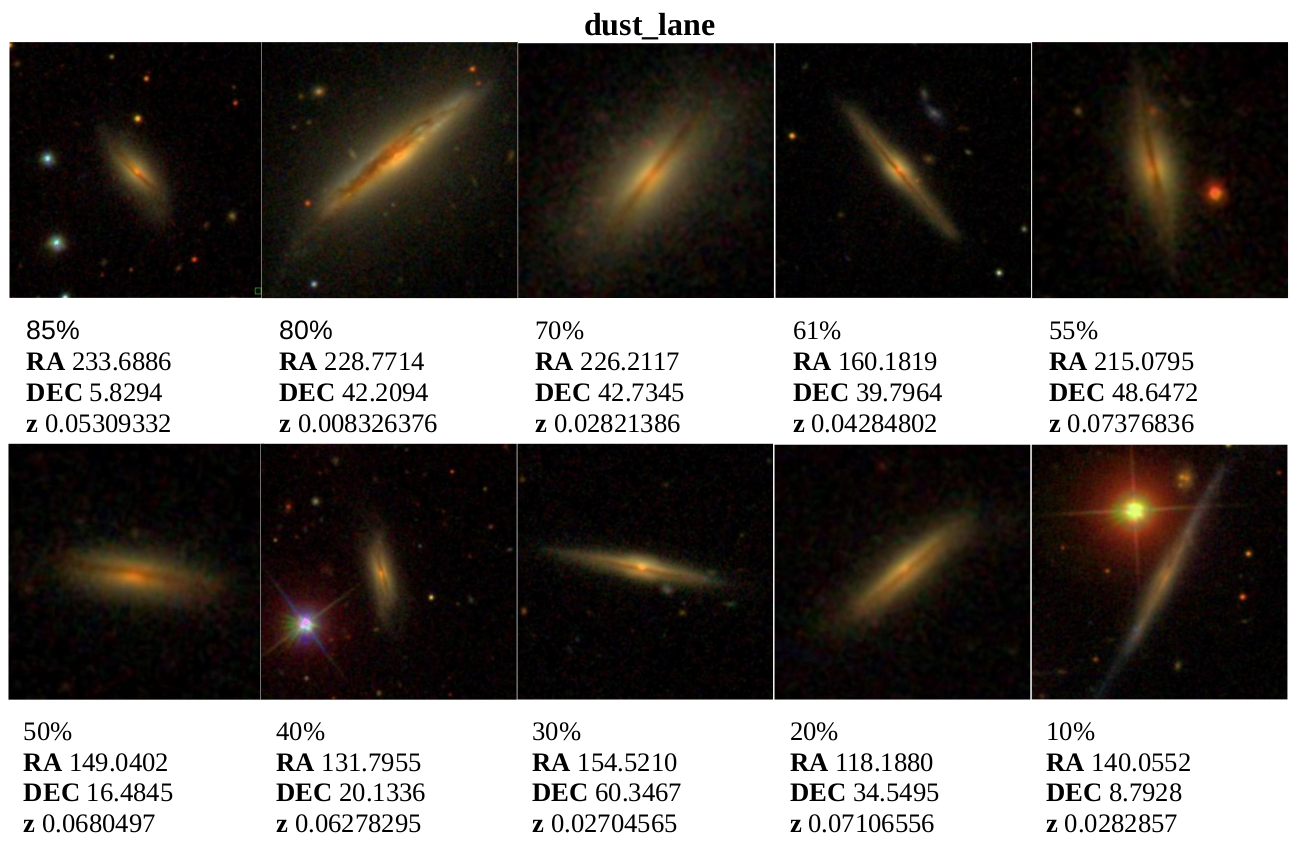}
     \caption{Examples of galaxies labeled as ``dust lane". In caption below each image: CNN probability to have this feature, RA and DEC, redshift.}
     \label{fig:dustlane}
 \end{figure*}

\begin{figure*}[!ht]
     \centering
     \includegraphics[scale=0.25]{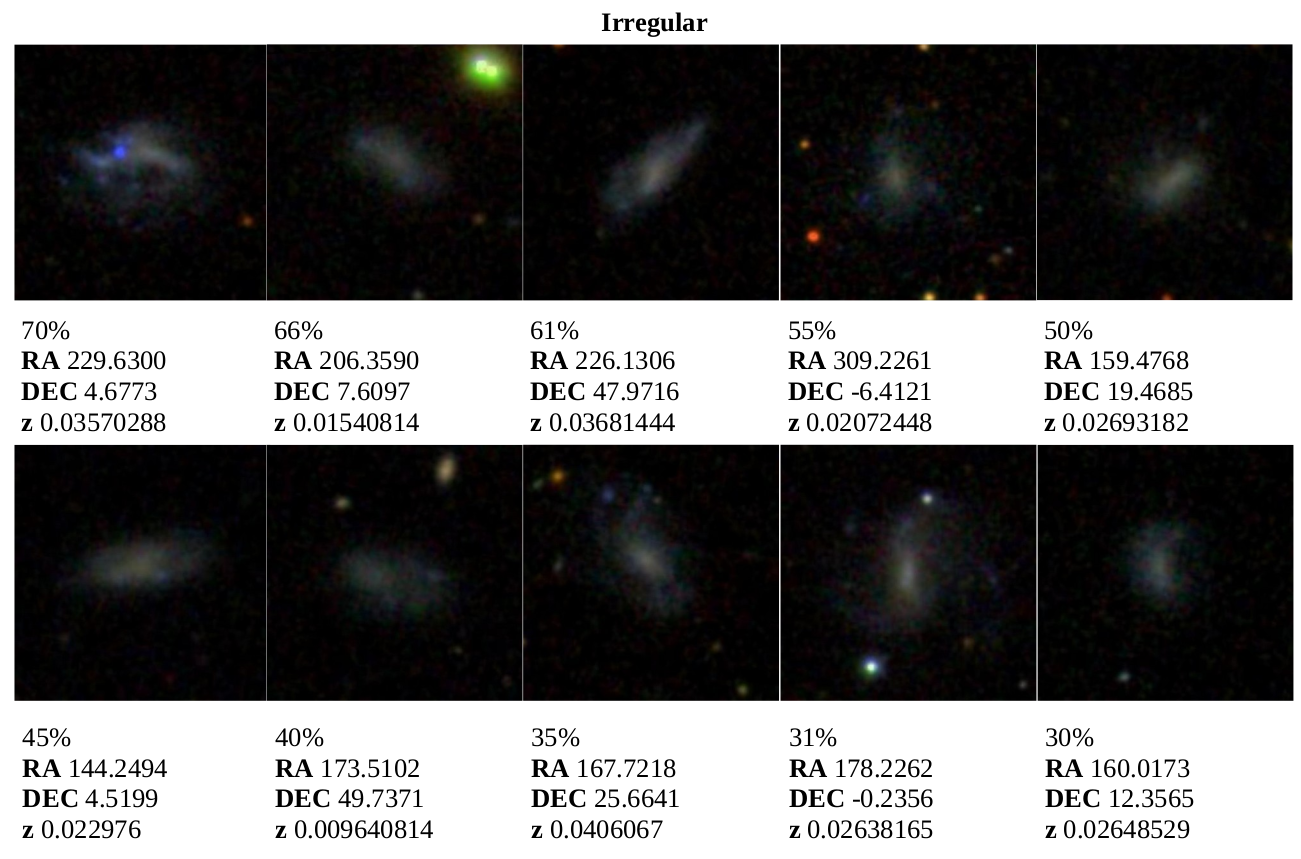}
     \caption{Examples of galaxies labeled as ``irregular". In caption below each image: CNN probability to have this feature, RA and DEC, redshift.}
     \label{fig:irregular}
 \end{figure*}

\begin{figure*}[!ht]
     \centering
     \includegraphics[scale=0.25]{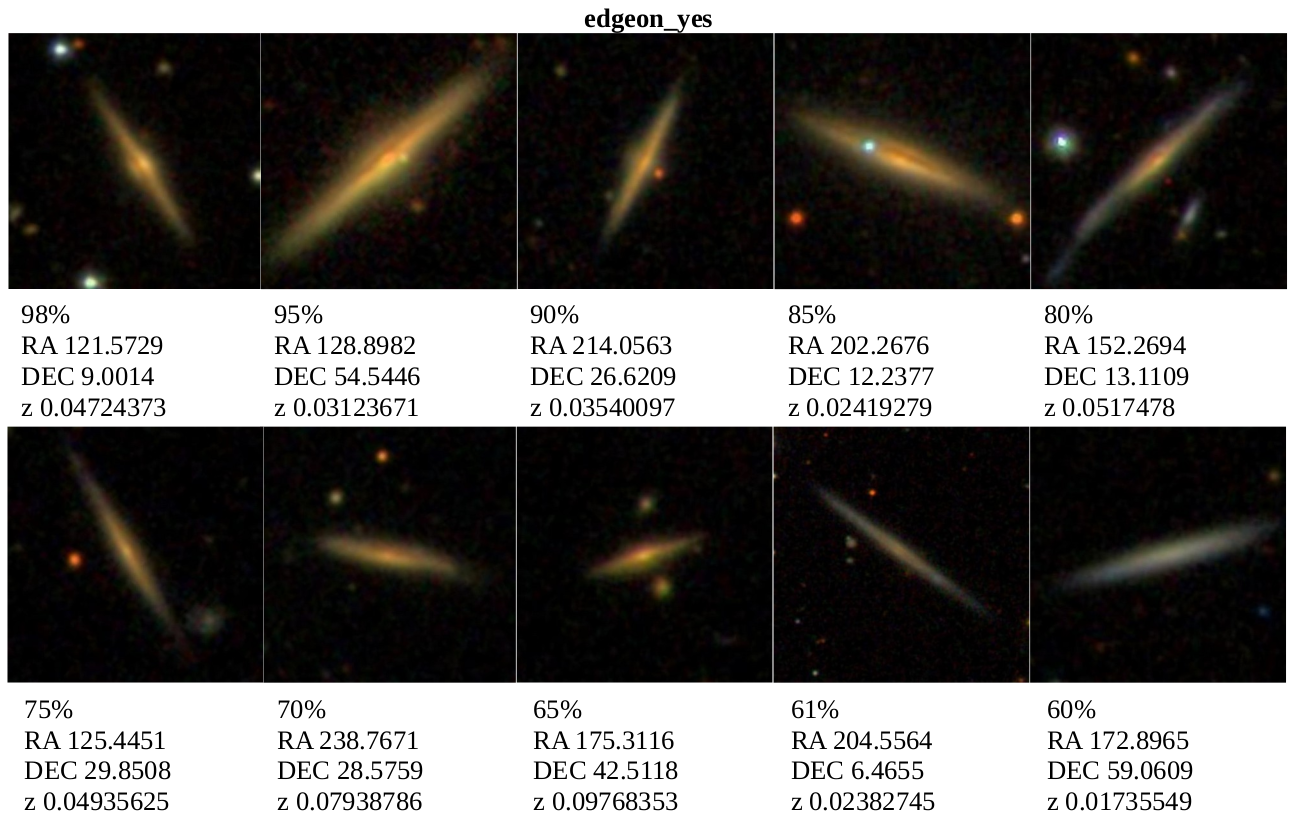}
     \caption{Examples of galaxies labeled as ``edge-on yes". In caption below each image: CNN probability to have this feature, RA and DEC, redshift.}
     \label{fig:edgeon}
 \end{figure*}

All the inspected galaxies labeled as ``dust lane", ``irregular", and ``edge-on yes"  demonstrate the perfect annotation. All these galaxies possess these features even having a lower probability by CNN classifier (see, examples in Fig. \ref{fig:dustlane}, \ref{fig:irregular}, \ref{fig:edgeon}: ``dust lane" in all range of probabilities, ``irregular" till 30\,\%, ``edge-on yes" till 60\,\%.

\subsection{Notes on problem points of CNN image-based galaxy classification by their features}

The evolutionary galaxy properties can affect ML methods' accuracy based on galaxies' photometry/image features. Among these misclassified types are the bluer HI-rich galaxies of early type and the redder HI-poor spiral galaxies; edge-on and galaxies seen face-on, especially with a pronounced bulge; the bulge-less (ultra-flat) galaxies with inclination $87^{\circ}\div90^{\circ}$ for seen edge-on and $10^\circ\div0^\circ$ for seen face-on. The face-on bulge-less galaxies can be considered as counterparts to the edge-on disk galaxies giving additional information on their physical parameters, including photometry \cite{Vavilova2021a, Smethurst2022}. So, their correct classification is very useful when compiling catalogs with a bulge to super-thin galaxies \cite{Kautsch2006, Bizyaev2014} or studying the influence of the environment on the morphology and quenching of galaxies in dense environments (for example, \cite{Lima2021} for the Hydra cluster). In such cases, where the surface brightness profile, color, and concentration indexes are needed, the ML algorithms trained over SDSS photometric parameters are less biased than when trained using GZ visual morphology, see, amongst others \cite{Dieleman2015, Dominguez2018, Dominguez2019, Vavilova2021a, Du2019}.

 At the same time, the results of applying the deep CNN to the images of our studied set \citep{Khramtsov2019a, Vasylenko2020} with the aim of binary morphological classification (late and early types) have shown limitations. Namely, DL methods can classify rounded galaxy images as ellipticals. Still, it cannot catch the SED properties of galaxies more clearly than the Support Vector Machine trained on the photometric features of galaxies. To avoid several of these misclassifications, Lingard et al. \cite{Lingard2020} developed a novel method, Galaxy Zoo Builder, working well with face-on galaxy image modeling based on the four-component photometric decomposition of spiral galaxies. Earlier,  Schawinski et al. \cite{Schawinski2014} exploited the SDSS, GALEX, and GZ data to substantiate the transformation from disk to elliptical morphology of low-redshift galaxies. 

Our visual inspection revealed yet several typical nuances of misclassified galaxy images.

\begin{figure*}[!ht]
     \centering
     \includegraphics[scale=0.25]{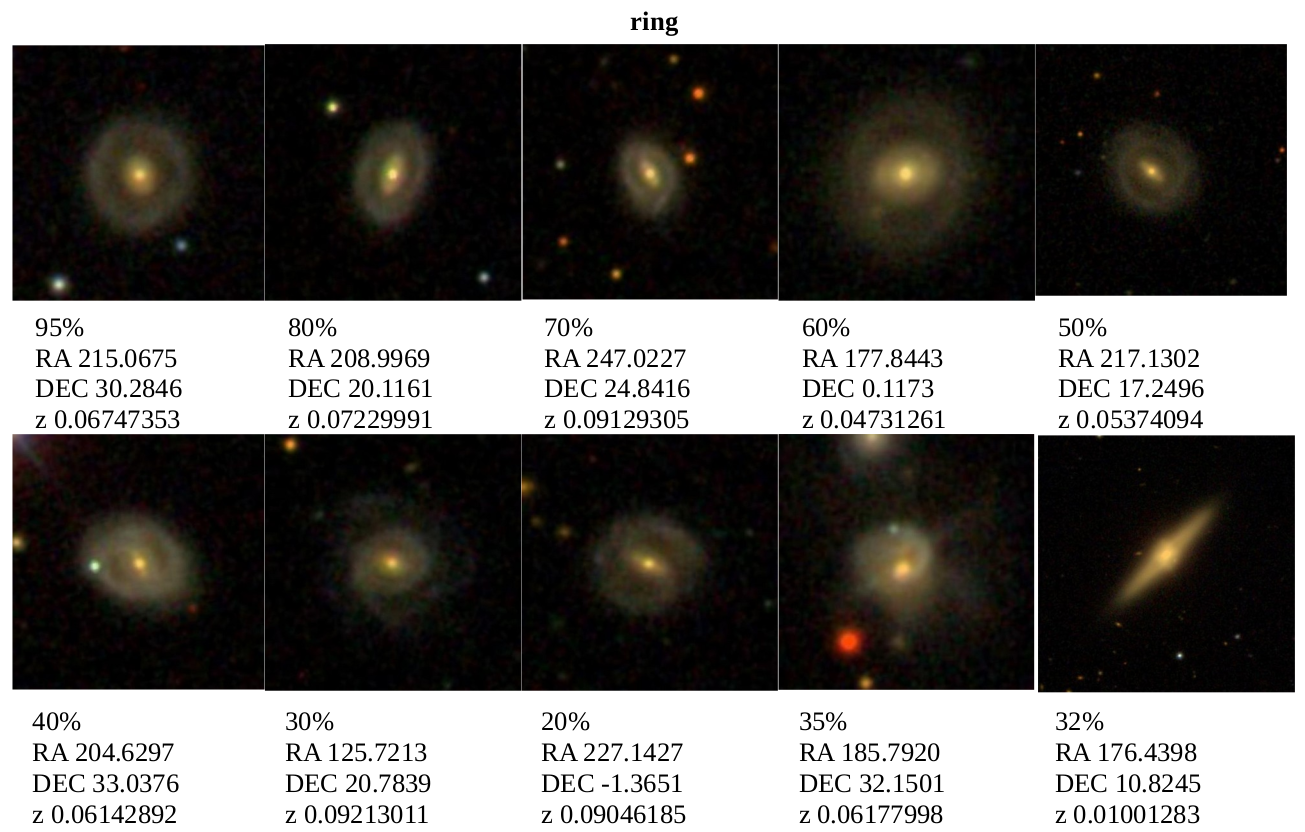}
     \caption{Examples of galaxies labeled as ``ring". In caption below each image: CNN probability to have this feature, RA and DEC, redshift.}
     \label{fig:ring}
 \end{figure*}

As related to the galaxies with ``ring" feature, we note that such galaxies were correctly labeled in all the range of probabilities. But there are misclassified images, mostly at the higher redshifts, which are a) the disk galaxies with bright bulge, b) galaxies with complicated contrast gradient of brightness (see, Fig. \ref{fig:ring}, two last images) as well as c) elliptical galaxies with a bright core, in which the brightness does not distributed smoothly towards the periphery, d) merging galaxies with a bright core and outer component distinctly differed in brightness, as a result, the neural network considers the outer component to be a ring. Creation of the representative catalogue of galaxies with ring(s) could be very useful. For example, Smirnov and Reshetnikov \cite{Smirnov2022} collected the samples of polar- and collision- ring galaxies from all the published data in several deep fields. Doing this painstaking preliminary search, they constructed the luminosity function for the ringed galaxies and confirmed the increase in their volume density with redshift: up to $z\sim1$ their density grows as $(1+z)^m$, where $m\geq5$. As related to problem point of elliptical galaxies with bright core, we link to the paper by Tarsitano et al. \cite{Tarsitano2022}, who developed a promising CNN approach based on the training of elliptical isophotes in the light distribution.

 \begin{figure*}[!ht]
     \centering
     \includegraphics[scale=0.35]{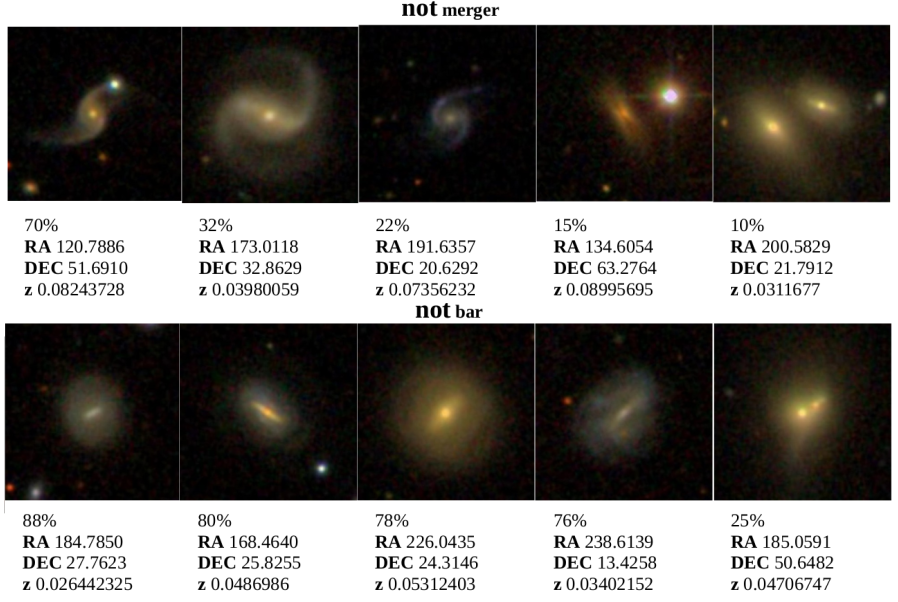}
     \caption{Examples of misclassified images of galaxies: with ``merger" and ``bar" features. In caption below each image: CNN probability to have this feature, RA and DEC, redshift.}
     \label{fig:misclass}
 \end{figure*}

The galaxy images labeled as with ``bar" have typical misclassifications. It has been happening when a) a central part of spiral arms of the edge-on galaxy is classified as a bar, b) the nuclei of merging galaxies are visually located near one another, then the CNN match this as a bar. The samples of misclassified images with ``bar" feature are in Fig. \ref{fig:misclass}. Bhambra et al. \cite{Bhambra2022} proposed the explainable artificial intelligence (XAI) techniques to measure galactic bar lengths and bulge-to-disk ratio. They used the Hoyle bar length catalogue \cite{Hoyle2011} vs. GZ annotated data and demonstrated that XAI works more successfully in predictions of a bar feature. Also, taking into account the class of ``smooth" galaxies (no bar, spiral arms, or other structure presents), these authors demonstrate the difficulties in reconciling differences between the ML model predictions and the GZ consensus. We agree with their conclusion that the citizen science method of classifying galaxies is less easily explained than ML methods.

We will not analyze the galaxy samples related to the ``spiral arms number" features. This task is perfectly studied by Hart et al. \cite{Hart2016}. Their method allowed to overcome where the rarer many-armed samples were incomplete, and the two-armed category suffered from sample contamination. They created a sample of about 18\,000 SDSS DR7 galaxies at $0.03<z<0.08$ with $M_{r}<-21$, which was sorted by arm multiplicity and further studied on star-forming activity. 

All the galaxy images with the ``star or artefacts" have these features. All of them contain galaxies that are classified. Bright stars and/or artifacts that obscure the image of a galaxy lead to misclassification of galaxies in most cases.

The sample of ``merger" galaxies also have false images, when a) galaxies are the optical pair, b) the star falls into the image background near a spiral galaxy, then the CNN considers the star as an elliptical galaxy and keeps it as merging, c) spiral galaxies without interaction, but their arms are untwisted (see, examples, in Fig. \ref{fig:misclass}. It is interesting to compare our results on merging galaxies with work by Reza \cite{Reza2021}, who also used the SDSS data and obtained that ExtraTrees classifier outperforms Neural Network for this distinct type of objects. It was noted that mergers are easily confused with both ellipticals and spirals when image-based classification is conducted.

As one can see, the CNN confident model predictions are highly accurate and allow us to filter big data collections of galaxy images with various morphological features. We expertized our obtained data and described several challenging images. When we develop the classification model, the aim is not only the state-of-the-art accuracy values but also defining problem points of the CNN model in working with galaxy images and training it to classify large surveys of galaxies no worse than an expert for small samples. 

\section{Conclusions}

The image-based CNN classifier was exploited by us to create a morphological catalog of 315\,776 SDSS DR9 low-redshift galaxies ($z<0.1$) following our previous works (\cite{Vavilova2021a, Vavilova2021b, Vavilova2022}). This target data set of the SDSS galaxies is tightly overlapped with the annotated data from GZ2 \citep{Willett2013}. For this reason, we divided it into two data sets: ``inference'', which do not match the GZ2 galaxies, and ``training'', which match the GZ2 galaxies. In the presence of a pronounced difference of visual parameters between galaxies from the GZ2 training data set and galaxies without known morphological parameters, we applied novel procedures, which allowed us to get rid of this difference, especially for smaller and fainter SDSS galaxies with $m_{r}<17.7$ from the inference data set. We describe in this paper how we applied the adversarial validation technique and managed the optimal train-test split of galaxies from the training data set to verify our CNN model based on the DenseNet-201 realistically. We have also found optimal galaxy image transformations, which help increase the classifier's generalization ability in similarity search, as is provided with a specifically created test data set.

We demonstrate for the first time that implication of the CNN model with train-test split of data sets and size-changing function simulating a decrease in magnitude and size (data augmentation) significantly improves the classification of smaller and fainter SDSS galaxies. It can be considered as another way to improve the human bias for those galaxy images that had a poor vote classification in the GZ project. Such an approach, likely auto-immunization, when the CNN classifier trained on very good images is able to retrain bad images from the same homogeneous sample, can be considered co-planar to other methods of combating the human bias.

The most interesting data products with this approach were obtained for galaxy classification by 34 detailed morphology features. The accuracy of CNN classifier is in the range of 83.3–99.4\,\% depending on 32 features (exception is for ``disturbed" (68.55\,\%) and ``arms winding medium" (77.39\,\%) features), the number of galaxies with the given feature in the inference data set, and the galaxy image quality (Table~\ref{tab:morphological_test1} and Table \ref{tab:morphological_test2}). To reach it, we calculated the number of galaxies that passed the best threshold for the acceptance of detailed morphological features. 

As a result, for the first time, we assigned the detailed morphological classification for more than 140\,000 low-redshift galaxies with $m_{r}<17.7$ from the SDSS DR9 (inference data set), which has the highest adversarial score by CNN classifier. The morphological catalogs of low-redshift SDSS galaxies with the most interesting features are available through the UkrVO web-site \url{http://ukr-vo.org/starcats/galaxies/} and will be supplemented to this paper through VizieR, as well as the catalog of galaxies with top five detailed morphological features (to wit, with a maximal prediction probability to posses such a feature). 

A visual inspection of the samples of galaxies with certain morphological features allowed to reveal typical problem points of galaxy image classification by shape and features from the astronomical point of view. We analyzed them in the discussion section, where we also compare machine learning photometry- and image- based approaches testifying that the best results are being performed with all of the galaxy data types (photometry, image, spectroscopy). We believe our results and notes on problem points will be useful to strength the CNN applicability and help in the morphological classification of galaxies within the current and forthcoming deep sky surveys at the peta-byte scale.

\section{Acknowledgements}
We thank Prof. Massimo Capacciolli and Dr. Valentina Karachentseva for the helpful discussion and remarks. The authors are deeply grateful to both referees for a detailed review of the article, their questions, comments and suggestions, which significantly improved the presentation of our results.

This paper uses data generated via the Zooniverse.org platformL development of which is funded by generous support including a Global Impact Award from Google and by a grant from the Alfred P. Sloan Foundation. This publication has been made possible by the participation of hundreds of thousands of volunteers in the Galaxy Zoo project. We thank the Galaxy Zoo team. The use of the SDSS \citep{Ahn2012, Blanton2017} and SAO/NASA Astrophysics Data System was extensively applicable. This study has also made with the NASA/IPAC Extragalactic Database (NED), which is operated by the Jet Propulsion Laboratory, California Institute of Technology, under contract with the NASA. This research has made use of the SIMBAD database, operated at CDS, Strasbourg, France \cite{Wenger2000}.

This work was done in frame of the Program of the NAS of Ukraine ``Support for the development of priority fields of scientific research” and the Target Program of Space Science of the NAS of Ukraine. Vavilova I.B. thanks the Wolfgang Pauli Institute, Vienna, Austria, for the support in frame of ``The Pauli Ukraine Project'' (2022) under the ``Models in plasma, Earth and space science'' program. 

%\end{acknowledgements}
%\section*{References}
\bibliography{arxiv-khramtsov.bib} % your references Yourfile.bib

\begin{thebibliography}{100}
\expandafter\ifx\csname url\endcsname\relax
  \def\url#1{\texttt{#1}}\fi
\expandafter\ifx\csname urlprefix\endcsname\relax\def\urlprefix{URL }\fi
\expandafter\ifx\csname href\endcsname\relax
  \def\href#1#2{#2} \def\path#1{#1}\fi

\bibitem{Agnello2015}
A.~{Agnello}, B.~C. {Kelly}, T.~{Treu}, P.~J. {Marshall}, {Data mining for
  gravitationally lensed quasars}, MNRAS 448~(2) (2015) 1446--1462.
\newblock \href {http://arxiv.org/abs/1410.4565} {\path{arXiv:1410.4565}},
  \href {http://dx.doi.org/10.1093/mnras/stv037}
  {\path{doi:10.1093/mnras/stv037}}.

\bibitem{Ostrovski2017}
F.~{Ostrovski}, R.~G. {McMahon}, A.~J. {Connolly}, C.~A. {Lemon}, {et al.},
  {VDES J2325-5229 a z = 2.7 gravitationally lensed quasar discovered using
  morphology-independent supervised machine learning}, MNRAS 465~(4) (2017)
  4325--4334.
\newblock \href {http://arxiv.org/abs/1607.01391} {\path{arXiv:1607.01391}},
  \href {http://dx.doi.org/10.1093/mnras/stw2958}
  {\path{doi:10.1093/mnras/stw2958}}.

\bibitem{Lanusse2018}
F.~{Lanusse}, Q.~{Ma}, N.~{Li}, T.~E. {Collett}, C.-L. {Li}, S.~{Ravanbakhsh},
  R.~{Mandelbaum}, B.~{P{\'o}czos}, {CMU DeepLens: deep learning for automatic
  image-based galaxy-galaxy strong lens finding}, MNRAS 473~(3) (2018)
  3895--3906.
\newblock \href {http://arxiv.org/abs/1703.02642} {\path{arXiv:1703.02642}},
  \href {http://dx.doi.org/10.1093/mnras/stx1665}
  {\path{doi:10.1093/mnras/stx1665}}.

\bibitem{Jacobs2019}
C.~{Jacobs}, T.~{Collett}, K.~{Glazebrook}, C.~{McCarthy}, {et al.}, {Finding
  high-redshift strong lenses in DES using convolutional neural networks},
  MNRAS 484~(4) (2019) 5330--5349.
\newblock \href {http://arxiv.org/abs/1811.03786} {\path{arXiv:1811.03786}},
  \href {http://dx.doi.org/10.1093/mnras/stz272}
  {\path{doi:10.1093/mnras/stz272}}.

\bibitem{Khramtsov2019b}
V.~{Khramtsov}, A.~{Sergeyev}, C.~{Spiniello}, C.~{Tortora}, {et al.},
  \href{https://doi.org/10.1051/0004-6361/201936006}{Kids-squad - ii. machine
  learning selection of bright extragalactic objects to search for new
  gravitationally lensed quasars}, A\&A 632 (2019) A56.
\newblock \href {http://dx.doi.org/10.1051/0004-6361/201936006}
  {\path{doi:10.1051/0004-6361/201936006}}.
\newline\urlprefix\url{https://doi.org/10.1051/0004-6361/201936006}

\bibitem{Petrillo2019}
C.~E. {Petrillo}, C.~{Tortora}, S.~{Chatterjee}, G.~{Vernardos}, L.~V.~E.
  {Koopmans}, G.~{Verdoes Kleijn}, N.~R. {Napolitano}, G.~{Covone}, L.~S.
  {Kelvin}, A.~M. {Hopkins}, {Testing convolutional neural networks for finding
  strong gravitational lenses in KiDS}, MNRAS 482~(1) (2019) 807--820.
\newblock \href {http://arxiv.org/abs/1807.04764} {\path{arXiv:1807.04764}},
  \href {http://dx.doi.org/10.1093/mnras/sty2683}
  {\path{doi:10.1093/mnras/sty2683}}.

\bibitem{Ribli2019}
D.~{Ribli}, B.~{\'A}. {Pataki}, J.~M. {Zorrilla Matilla}, D.~{Hsu},
  Z.~{Haiman}, I.~{Csabai}, {Weak lensing cosmology with convolutional neural
  networks on noisy data}, MNRAS 490~(2) (2019) 1843--1860.
\newblock \href {http://arxiv.org/abs/1902.03663} {\path{arXiv:1902.03663}},
  \href {http://dx.doi.org/10.1093/mnras/stz2610}
  {\path{doi:10.1093/mnras/stz2610}}.

\bibitem{Pourrahmani2018}
M.~{Pourrahmani}, H.~{Nayyeri}, A.~{Cooray}, {LensFlow: A Convolutional Neural
  Network in Search of Strong Gravitational Lenses}, ApJ 856~(1) (2018) 68.
\newblock \href {http://arxiv.org/abs/1705.05857} {\path{arXiv:1705.05857}},
  \href {http://dx.doi.org/10.3847/1538-4357/aaae6a}
  {\path{doi:10.3847/1538-4357/aaae6a}}.

\bibitem{Pasquet2019}
J.~{Pasquet}, E.~{Bertin}, M.~{Treyer}, S.~{Arnouts}, D.~{Fouchez},
  {Photometric redshifts from SDSS images using a convolutional neural
  network}, A\&A 621 (2019) A26.
\newblock \href {http://arxiv.org/abs/1806.06607} {\path{arXiv:1806.06607}},
  \href {http://dx.doi.org/10.1051/0004-6361/201833617}
  {\path{doi:10.1051/0004-6361/201833617}}.

\bibitem{Fussell2019}
L.~{Fussell}, B.~{Moews}, {Forging new worlds: high-resolution synthetic
  galaxies with chained generative adversarial networks}, MNRAS 485~(3) (2019)
  3203--3214.
\newblock \href {http://arxiv.org/abs/1811.03081} {\path{arXiv:1811.03081}},
  \href {http://dx.doi.org/10.1093/mnras/stz602}
  {\path{doi:10.1093/mnras/stz602}}.

\bibitem{Salvato2019}
M.~{Salvato}, O.~{Ilbert}, B.~{Hoyle}, {The many flavours of photometric
  redshifts}, Nature Astronomy 3 (2019) 212--222.
\newblock \href {http://arxiv.org/abs/1805.12574} {\path{arXiv:1805.12574}},
  \href {http://dx.doi.org/10.1038/s41550-018-0478-0}
  {\path{doi:10.1038/s41550-018-0478-0}}.

\bibitem{Bonnett2016}
C.~{Bonnett}, M.~A. {Troxel}, W.~{Hartley}, A.~{Amara}, {et al.}, {Redshift
  distributions of galaxies in the Dark Energy Survey Science Verification
  shear catalogue and implications for weak lensing}, PRD 94~(4) (2016) 042005.
\newblock \href {http://arxiv.org/abs/1507.05909} {\path{arXiv:1507.05909}},
  \href {http://dx.doi.org/10.1103/PhysRevD.94.042005}
  {\path{doi:10.1103/PhysRevD.94.042005}}.

\bibitem{Amaro2019}
V.~{Amaro}, S.~{Cavuoti}, M.~{Brescia}, C.~{Vellucci}, {et al.}, {Statistical
  analysis of probability density functions for photometric redshifts through
  the KiDS-ESO-DR3 galaxies}, MNRAS 482~(3) (2019) 3116--3134.
\newblock \href {http://arxiv.org/abs/1810.09777} {\path{arXiv:1810.09777}},
  \href {http://dx.doi.org/10.1093/mnras/sty2922}
  {\path{doi:10.1093/mnras/sty2922}}.

\bibitem{Sadeh2016}
I.~{Sadeh}, F.~B. {Abdalla}, O.~{Lahav}, {ANNz2: Photometric Redshift and
  Probability Distribution Function Estimation using Machine Learning}, PASP
  128~(968) (2016) 104502.
\newblock \href {http://arxiv.org/abs/1507.00490} {\path{arXiv:1507.00490}},
  \href {http://dx.doi.org/10.1088/1538-3873/128/968/104502}
  {\path{doi:10.1088/1538-3873/128/968/104502}}.

\bibitem{Pasquet2018}
J.~{Pasquet-Itam}, J.~{Pasquet}, {Deep learning approach for classifying,
  detecting and predicting photometric redshifts of quasars in the Sloan
  Digital Sky Survey stripe 82}, A\&A 611 (2018) A97.
\newblock \href {http://arxiv.org/abs/1712.02777} {\path{arXiv:1712.02777}},
  \href {http://dx.doi.org/10.1051/0004-6361/201731106}
  {\path{doi:10.1051/0004-6361/201731106}}.

\bibitem{Kugler2016}
S.~D. {K{\"u}gler}, N.~{Gianniotis}, {Modelling multimodal photometric redshift
  regression with noisy observations}, arXiv e-prints\href
  {http://arxiv.org/abs/1607.06059} {\path{arXiv:1607.06059}}.

\bibitem{Speagle2017}
J.~S. {Speagle}, D.~J. {Eisenstein}, {Deriving photometric redshifts using
  fuzzy archetypes and self-organizing maps - II. Implementation}, MNRAS
  469~(1) (2017) 1205--1224.
\newblock \href {http://dx.doi.org/10.1093/mnras/stx510}
  {\path{doi:10.1093/mnras/stx510}}.

\bibitem{Disanto2018}
A.~{D'Isanto}, S.~{Cavuoti}, F.~{Gieseke}, K.~L. {Polsterer}, {Return of the
  features. Efficient feature selection and interpretation for photometric
  redshifts}, A\&A 616 (2018) A97.
\newblock \href {http://arxiv.org/abs/1803.10032} {\path{arXiv:1803.10032}},
  \href {http://dx.doi.org/10.1051/0004-6361/201833103}
  {\path{doi:10.1051/0004-6361/201833103}}.

\bibitem{Elyiv2020}
A.~A. {Elyiv}, O.~V. {Melnyk}, I.~B. {Vavilova}, D.~V. {Dobrycheva}, V.~E.
  {Karachentseva}, {Machine-learning computation of distance modulus for local
  galaxies}, A\&A 635 (2020) A124.
\newblock \href {http://arxiv.org/abs/1910.07317} {\path{arXiv:1910.07317}},
  \href {http://dx.doi.org/10.1051/0004-6361/201936883}
  {\path{doi:10.1051/0004-6361/201936883}}.

\bibitem{Rastegarnia2022}
F.~{Rastegarnia}, M.~T. {Mirtorabi}, R.~{Moradi}, A.~{Vafaei Sadr}, Y.~{Wang},
  {Deep learning in searching the spectroscopic redshift of quasars}, MNRAS
  511~(3) (2022) 4490--4499.
\newblock \href {http://arxiv.org/abs/2201.03393} {\path{arXiv:2201.03393}},
  \href {http://dx.doi.org/10.1093/mnras/stac076}
  {\path{doi:10.1093/mnras/stac076}}.

\bibitem{Schawinski2017}
K.~{Schawinski}, C.~{Zhang}, H.~{Zhang}, L.~{Fowler}, G.~K. {Santhanam},
  {Generative adversarial networks recover features in astrophysical images of
  galaxies beyond the deconvolution limit}, MNRAS 467~(1) (2017) L110--L114.
\newblock \href {http://arxiv.org/abs/1702.00403} {\path{arXiv:1702.00403}},
  \href {http://dx.doi.org/10.1093/mnrasl/slx008}
  {\path{doi:10.1093/mnrasl/slx008}}.

\bibitem{Vavilova2018}
I.~B. {Vavilova}, A.~A. {Elyiv}, M.~Y. {Vasylenko}, {Behind the Zone of
  Avoidance of the Milky Way: what can we Restore by Direct and Indirect
  Methods?}, Russian Radio Physics and Radio Astronomy 23~(4) (2018) 244--257.
\newblock \href {http://dx.doi.org/10.15407/rpra23.04.244}
  {\path{doi:10.15407/rpra23.04.244}}.

\bibitem{Diakogiannis2019}
F.~I. {Diakogiannis}, G.~F. {Lewis}, R.~A. {Ibata}, M.~{Guglielmo}, M.~I.
  {Wilkinson}, C.~{Power}, {Reliable mass calculation in spherical gravitating
  systems}, MNRAS 482~(3) (2019) 3356--3372.
\newblock \href {http://arxiv.org/abs/1810.11375} {\path{arXiv:1810.11375}},
  \href {http://dx.doi.org/10.1093/mnras/sty2931}
  {\path{doi:10.1093/mnras/sty2931}}.

\bibitem{Tsizh2020}
M.~{Tsizh}, B.~{Novosyadlyj}, Y.~{Holovatch}, N.~I. {Libeskind}, {Large-scale
  structures in the {\ensuremath{\Lambda}}CDM Universe: network analysis and
  machine learning}, MNRAS 495~(1) (2020) 1311--1320.
\newblock \href {http://arxiv.org/abs/1910.07868} {\path{arXiv:1910.07868}},
  \href {http://dx.doi.org/10.1093/mnras/staa1030}
  {\path{doi:10.1093/mnras/staa1030}}.

\bibitem{Chen2020}
Y.~{Chen}, H.~J. {Mo}, C.~{Li}, H.~{Wang}, X.~{Yang}, Y.~{Zhang}, K.~{Wang},
  {Relating the Structure of Dark Matter Halos to Their Assembly and
  Environment}, Astrophys.J. 899~(1) (2020) 81.
\newblock \href {http://arxiv.org/abs/2003.05137} {\path{arXiv:2003.05137}},
  \href {http://dx.doi.org/10.3847/1538-4357/aba597}
  {\path{doi:10.3847/1538-4357/aba597}}.

\bibitem{Moriwaki2021}
K.~{Moriwaki}, M.~{Shirasaki}, N.~{Yoshida}, {Deep Learning for Line Intensity
  Mapping Observations: Information Extraction from Noisy Maps}, ApJL 906~(1)
  (2021) L1.
\newblock \href {http://arxiv.org/abs/2010.00809} {\path{arXiv:2010.00809}},
  \href {http://dx.doi.org/10.3847/2041-8213/abd17f}
  {\path{doi:10.3847/2041-8213/abd17f}}.

\bibitem{Flamary2016}
R.~{Flamary}, {Astronomical image reconstruction with convolutional neural
  networks}, arXiv e-prints\href {http://arxiv.org/abs/1612.04526}
  {\path{arXiv:1612.04526}}.

\bibitem{Kremer2017}
J.~{Kremer}, K.~{Stensbo-Smidt}, F.~{Gieseke}, K.~{Steenstrup Pedersen},
  C.~{Igel}, {Big Universe, Big Data: Machine Learning and Image Analysis for
  Astronomy}, arXiv e-prints (2017) arXiv:1704.04650\href
  {http://arxiv.org/abs/1704.04650} {\path{arXiv:1704.04650}}.

\bibitem{Reiman2019}
D.~M. {Reiman}, B.~E. {G{\"o}hre}, {Deblending galaxy superpositions with
  branched generative adversarial networks}, MNRAS 485~(2) (2019) 2617--2627.
\newblock \href {http://arxiv.org/abs/1810.10098} {\path{arXiv:1810.10098}},
  \href {http://dx.doi.org/10.1093/mnras/stz575}
  {\path{doi:10.1093/mnras/stz575}}.

\bibitem{Buchanan2021}
J.~J. {Buchanan}, M.~D. {Schneider}, R.~E. {Armstrong}, A.~L. {Muyskens}, B.~W.
  {Priest}, R.~J. {Dana}, {Gaussian Process Classification for Galaxy Blend
  Identification in LSST}, arXiv e-prints (2021) arXiv:2107.09246\href
  {http://arxiv.org/abs/2107.09246} {\path{arXiv:2107.09246}}.

\bibitem{Bouchefry2020}
K.~{El Bouchefry}, R.~S. {de Souza}, {Learning in Big Data: Introduction to
  Machine Learning}, in: P.~{{\v{S}}koda}, F.~{Adam} (Eds.), Knowledge
  Discovery in Big Data from Astronomy and Earth Observation, 2020, pp.
  225--249.
\newblock \href {http://dx.doi.org/10.1016/B978-0-12-819154-5.00023-0}
  {\path{doi:10.1016/B978-0-12-819154-5.00023-0}}.

\bibitem{Burgazli2022}
A.~{Burgazli}, O.~{Sergijenko}, I.~{Vavilova}, {Machine learning in cosmology
  and gravitational wave astronomy: recent trends}, Vol.~22, In: Horizons in
  Computer Science Research, Ed. T.S. Claryvol, New York, Nova Science
  Publisher Inc., 2022, Ch.~7, pp. 193--240.

\bibitem{Kang2019}
S.-J. {Kang}, J.-H. {Fan}, W.~{Mao}, Q.~{Wu}, J.~{Feng}, Y.~{Yin}, {Evaluating
  the Optical Classification of Fermi BCUs Using Machine Learning}, ApJ 872~(2)
  (2019) 189.
\newblock \href {http://arxiv.org/abs/1902.07717} {\path{arXiv:1902.07717}},
  \href {http://dx.doi.org/10.3847/1538-4357/ab0383}
  {\path{doi:10.3847/1538-4357/ab0383}}.

\bibitem{Krause2017}
M.~{Krause}, E.~{Pueschel}, G.~{Maier}, {Improved {\ensuremath{\gamma}}/hadron
  separation for the detection of faint {\ensuremath{\gamma}}-ray sources using
  boosted decision trees}, Astroparticle Physics 89 (2017) 1--9.
\newblock \href {http://arxiv.org/abs/1701.06928} {\path{arXiv:1701.06928}},
  \href {http://dx.doi.org/10.1016/j.astropartphys.2017.01.004}
  {\path{doi:10.1016/j.astropartphys.2017.01.004}}.

\bibitem{Ruhe2020}
T.~{Ruhe}, {Application of machine learning algorithms in imaging Cherenkov and
  neutrino astronomy}, International Journal of Modern Physics A 35~(33) (2020)
  2043004--778.
\newblock \href {http://dx.doi.org/10.1142/S0217751X20430046}
  {\path{doi:10.1142/S0217751X20430046}}.

\bibitem{Morello2018}
G.~{Morello}, P.~W. {Morris}, S.~D. {Van Dyk}, A.~P. {Marston}, J.~C.
  {Mauerhan}, {Applications of machine-learning algorithms for infrared colour
  selection of Galactic Wolf-Rayet stars}, MNRAS 473~(2) (2018) 2565--2574.
\newblock \href {http://arxiv.org/abs/1712.01409} {\path{arXiv:1712.01409}},
  \href {http://dx.doi.org/10.1093/mnras/stx2474}
  {\path{doi:10.1093/mnras/stx2474}}.

\bibitem{Ciuca2017}
R.~{Ciuca}, O.~F. {Hern{\'a}ndez}, {A Bayesian framework for cosmic string
  searches in CMB maps}, J. Cosm. Astropart. Phys. 2017~(8) (2017) 028.
\newblock \href {http://arxiv.org/abs/1706.04131} {\path{arXiv:1706.04131}},
  \href {http://dx.doi.org/10.1088/1475-7516/2017/08/028}
  {\path{doi:10.1088/1475-7516/2017/08/028}}.

\bibitem{Aniyan2017}
A.~K. {Aniyan}, K.~{Thorat}, {Classifying Radio Galaxies with the Convolutional
  Neural Network}, ApJs 230~(2) (2017) 20.
\newblock \href {http://arxiv.org/abs/1705.03413} {\path{arXiv:1705.03413}},
  \href {http://dx.doi.org/10.3847/1538-4365/aa7333}
  {\path{doi:10.3847/1538-4365/aa7333}}.

\bibitem{Lukic2018}
V.~{Lukic}, M.~{Br{\"u}ggen}, J.~K. {Banfield}, O.~I. {Wong}, L.~{Rudnick},
  R.~P. {Norris}, B.~{Simmons}, {Radio Galaxy Zoo: compact and extended radio
  source classification with deep learning}, MNRAS 476~(1) (2018) 246--260.
\newblock \href {http://arxiv.org/abs/1801.04861} {\path{arXiv:1801.04861}},
  \href {http://dx.doi.org/10.1093/mnras/sty163}
  {\path{doi:10.1093/mnras/sty163}}.

\bibitem{Ma2019}
Z.~{Ma}, H.~{Xu}, J.~{Zhu}, D.~{Hu}, {et al.}, {A Machine Learning Based
  Morphological Classification of 14,245 Radio AGNs Selected from the
  Best-Heckman Sample}, ApJS 240~(2) (2019) 34.
\newblock \href {http://arxiv.org/abs/1812.07190} {\path{arXiv:1812.07190}},
  \href {http://dx.doi.org/10.3847/1538-4365/aaf9a2}
  {\path{doi:10.3847/1538-4365/aaf9a2}}.

\bibitem{Scaife2021}
A.~M.~M. {Scaife}, F.~{Porter}, {Fanaroff-Riley classification of radio
  galaxies using group-equivariant convolutional neural networks}, MNRAS
  503~(2) (2021) 2369--2379.
\newblock \href {http://arxiv.org/abs/2102.08252} {\path{arXiv:2102.08252}},
  \href {http://dx.doi.org/10.1093/mnras/stab530}
  {\path{doi:10.1093/mnras/stab530}}.

\bibitem{Ciprijanovic2021}
A.~{{\'C}iprijanovi{\'c}}, D.~{Kafkes}, K.~{Downey}, S.~{Jenkins}, {et al.},
  {DeepMerge - II. Building robust deep learning algorithms for merging galaxy
  identification across domains}, MNRAS 506~(1) (2021) 677--691.
\newblock \href {http://arxiv.org/abs/2103.01373} {\path{arXiv:2103.01373}},
  \href {http://dx.doi.org/10.1093/mnras/stab1677}
  {\path{doi:10.1093/mnras/stab1677}}.

\bibitem{Shamir2021}
L.~{Shamir}, {Automatic identification of outliers in Hubble Space Telescope
  galaxy images}, MNRAS 501~(4) (2021) 5229--5238.
\newblock \href {http://arxiv.org/abs/2101.02623} {\path{arXiv:2101.02623}},
  \href {http://dx.doi.org/10.1093/mnras/staa4036}
  {\path{doi:10.1093/mnras/staa4036}}.

\bibitem{Vavilova2021a}
I.~B. {Vavilova}, D.~V. {Dobrycheva}, M.~Y. {Vasylenko}, A.~A. {Elyiv}, O.~V.
  {Melnyk}, V.~{Khramtsov}, {Machine learning technique for morphological
  classification of galaxies from the SDSS. I. Photometry-based approach}, A\&A
  648 (2021) A122.
\newblock \href {http://arxiv.org/abs/1712.08955} {\path{arXiv:1712.08955}},
  \href {http://dx.doi.org/10.1051/0004-6361/202038981}
  {\path{doi:10.1051/0004-6361/202038981}}.

\bibitem{Vavilova2022}
I.~B. {Vavilova}, V.~{Khramtsov}, D.~V. {Dobrycheva}, M.~Y. {Vasylenko}, A.~A.
  {Elyiv}, O.~V. {Melnyk}, {Machine learning technique for morphological
  classification of galaxies from SDSS. II. The image-based morphological
  catalogs of galaxies at 0.02<z<0.1}, Space Sci. \& Technol. (2022) 3--22\href
  {http://arxiv.org/abs/2203.06373} {\path{arXiv:2203.06373}}, \href
  {http://dx.doi.org/doi.org/10.15407/knit2022.01.003}
  {\path{doi:doi.org/10.15407/knit2022.01.003}}.

\bibitem{Walmsley2020}
M.~{Walmsley}, L.~{Smith}, C.~{Lintott}, Y.~{Gal}, {et al.}, {Galaxy Zoo:
  probabilistic morphology through Bayesian CNNs and active learning}, MNRAS
  491~(2) (2020) 1554--1574.
\newblock \href {http://arxiv.org/abs/1905.07424} {\path{arXiv:1905.07424}},
  \href {http://dx.doi.org/10.1093/mnras/stz2816}
  {\path{doi:10.1093/mnras/stz2816}}.

\bibitem{Muller2016}
A.~{Muller}, S.~{Guido}, Introduction to Machine Learning with Python,
  O’Reilly Media, 2016.

\bibitem{Melnyk2012}
O.~V. {Melnyk}, D.~V. {Dobrycheva}, I.~B. {Vavilova}, {Morphology and color
  indices of galaxies in Pairs: Criteria for the classification of galaxies},
  Astrophysics 55~(3) (2012) 293--305.
\newblock \href {http://dx.doi.org/10.1007/s10511-012-9236-7}
  {\path{doi:10.1007/s10511-012-9236-7}}.

\bibitem{Dobrycheva2014}
D.~V. {Dobrycheva}, O.~V. {Melnyk}, I.~B. {Vavilova}, A.~A. {Elyiv},
  {Environmental Properties of Galaxies at z < 0.1 from the SDSS via the
  Voronoi Tessellation}, Odessa Astronomical Publications 27 (2014) 26.

\bibitem{Dobrycheva2015}
D.~V. {Dobrycheva}, O.~V. {Melnyk}, I.~B. {Vavilova}, A.~A. {Elyiv},
  {Environmental Density vs. Colour Indices of the Low Redshifts Galaxies},
  Astrophysics 58~(2) (2015) 168--180.
\newblock \href {http://dx.doi.org/10.1007/s10511-015-9373-x}
  {\path{doi:10.1007/s10511-015-9373-x}}.

\bibitem{Dobrycheva2017}
D.~V. {Dobrycheva}, I.~B. {Vavilova}, O.~V. {Melnyk}, A.~A. {Elyiv}, {Machine
  learning technique for morphological classification of galaxies at z < 0.1
  from the SDSS}, arXiv e-prints\href {http://arxiv.org/abs/1712.08955}
  {\path{arXiv:1712.08955}}.

\bibitem{Dobrycheva2017a}
D.~V. {Dobrycheva}, {Morphological content and color indices bimodality of a
  new galaxy sample at the redshifts z < 0.1}, Ph.D. thesis, Main Astronomical
  Observatory, NAS of Ukraine (Apr. 2017).

\bibitem{Dobrycheva2018}
D.~V. {Dobrycheva}, I.~B. {Vavilova}, O.~V. {Melnyk}, A.~A. {Elyiv},
  {Morphological Type and Color Indices of the SDSS DR9 Galaxies at 0.02 \&lt;
  z {\ensuremath{\leq}} 0.06}, Kinematics and Physics of Celestial Bodies
  34~(6) (2018) 290--301.
\newblock \href {http://dx.doi.org/10.3103/S0884591318060028}
  {\path{doi:10.3103/S0884591318060028}}.

\bibitem{Vasylenko2019}
M.~Y. {Vasylenko}, D.~V. {Dobrycheva}, I.~B. {Vavilova}, O.~V. {Melnyk}, A.~A.
  {Elyiv}, {Verification of Machine Learning Methods for Binary Morphological
  Classification of Galaxies from SDSS}, Odessa Astronomical Publications 32
  (2019) 46.
\newblock \href {http://dx.doi.org/10.18524/1810-4215.2019.32.182538}
  {\path{doi:10.18524/1810-4215.2019.32.182538}}.

\bibitem{Khramtsov2019a}
V.~{Khramtsov}, D.~V. {Dobrycheva}, M.~Y. {Vasylenko}, V.~S. {Akhmetov},
  \href{http://oap.onu.edu.ua/article/view/182092}{Deep learning for
  morphological classification of galaxies from sdss}, Odessa Astronomical
  Publications 32 (2019) 21.
\newblock \href {http://dx.doi.org/10.18524/1810-4215.2019.32.182092}
  {\path{doi:10.18524/1810-4215.2019.32.182092}}.
\newline\urlprefix\url{http://oap.onu.edu.ua/article/view/182092}

\bibitem{Vasylenko2020}
M.~{Vasylenko}, D.~{Dobrycheva}, V.~{Khramtsov}, I.~{Vavilova}, {Deep
  Convolutional Neural Networks models for the binary morphological
  classification of SDSS-galaxies}, Communications of the Byurakan
  Astrophysical Observatory 67 (2020) 354.
\newblock \href {http://dx.doi.org/10.52526/25792776-2020.67.2-354}
  {\path{doi:10.52526/25792776-2020.67.2-354}}.

\bibitem{Vavilova2020a}
I.~{Vavilova}, D.~{Dobrycheva}, M.~{Vasylenko}, A.~{Elyiv}, O.~{Melnyk},
  {Multiwavelength Extragalactic Surveys: Examples of Data Mining}, In:
  Knowledge Discovery in Big Data from Astronomy and Earth Observation, Eds. P.
  Skoda and F. Adam, Elsevier, 2020, Ch.~16, pp. 307--323.
\newblock \href {http://dx.doi.org/10.1016/B978-0-12-819154-5.00028-X}
  {\path{doi:10.1016/B978-0-12-819154-5.00028-X}}.

\bibitem{Vavilova2021c}
I.~{Vavilova}, A.~{Elyiv}, D.~{Dobrycheva}, O.~{Melnyk}, {The Voronoi
  tessellation method in astronomy}, Vol.~39, In: Intelligent Astrophysics,
  Eds. I. Zelinka, M. Brescia, D. Baron, Springer, Cham, 2021, Ch.~3, pp.
  57--79.
\newblock \href {http://dx.doi.org/10.1007/978-3-030-65867-0\_3}
  {\path{doi:10.1007/978-3-030-65867-0\_3}}.

\bibitem{Vavilova2021b}
I.~B. {Vavilova}, D.~V. {Dobrycheva}, M.~Y. {Vasylenko}, A.~A. {Elyiv}, O.~V.
  {Melnyk}, V.~{Khramtsov}, {VizieR Online Data Catalog: SDSS galaxies
  morphological classification (Vavilova+, 2021)}, VizieR Online Data Catalog
  (2021) J/A+A/648/A122.

\bibitem{Vavilova2022b}
I.~B. {Vavilova}, V.~{Khramtsov}, D.~V. {Dobrycheva}, M.~Y. {Vasylenko}, A.~A.
  {Elyiv}, O.~V. {Melnyk}, {VizieR Online Data Catalog: Galaxies at 0.02<z<0.1
  morphological catalog (Vavilova+, 2022)}, VizieR Online Data Catalog (2022)
  J/other/KNIT/28.3/gal5mcls.

\bibitem{Willett2013}
K.~W. {Willett}, C.~J. {Lintott}, S.~P. {Bamford}, {et al.}, {Galaxy Zoo 2:
  detailed morphological classifications for 304 122 galaxies from the Sloan
  Digital Sky Survey}, MNRAS 435~(4) (2013) 2835--2860.
\newblock \href {http://arxiv.org/abs/1308.3496} {\path{arXiv:1308.3496}},
  \href {http://dx.doi.org/10.1093/mnras/stt1458}
  {\path{doi:10.1093/mnras/stt1458}}.

\bibitem{Blanton2001}
M.~R. {Blanton}, J.~{Dalcanton}, D.~{Eisenstein}, J.~{Loveday}, {et al.}, {The
  Luminosity Function of Galaxies in SDSS Commissioning Data}, AJ 121~(5)
  (2001) 2358--2380.
\newblock \href {http://arxiv.org/abs/astro-ph/0012085}
  {\path{arXiv:astro-ph/0012085}}, \href {http://dx.doi.org/10.1086/320405}
  {\path{doi:10.1086/320405}}.

\bibitem{Yasuda2001}
N.~{Yasuda}, M.~{Fukugita}, V.~K. {Narayanan}, R.~H. {Lupton}, {et al.},
  {Galaxy Number Counts from the Sloan Digital Sky Survey Commissioning Data},
  AJ 122~(3) (2001) 1104--1124.
\newblock \href {http://arxiv.org/abs/astro-ph/0105545}
  {\path{arXiv:astro-ph/0105545}}, \href {http://dx.doi.org/10.1086/322093}
  {\path{doi:10.1086/322093}}.

\bibitem{walmsley2021galaxy}
M.~Walmsley, C.~Lintott, T.~Geron, S.~Kruk, C.~Krawczyk, K.~W. Willett,
  S.~Bamford, W.~Keel, L.~S. Kelvin, L.~Fortson, K.~L. Masters, V.~Mehta, B.~D.
  Simmons, R.~Smethurst, E.~M. Baeten, C.~Macmillan, Galaxy zoo decals:
  Detailed visual morphology measurements from volunteers and deep learning for
  314000 galaxies (2021).
\newblock \href {http://arxiv.org/abs/2102.08414} {\path{arXiv:2102.08414}}.

\bibitem{Lupton2004}
R.~{Lupton}, M.~R. {Blanton}, G.~{Fekete}, D.~W. {Hogg}, W.~{O'Mullane},
  A.~{Szalay}, N.~{Wherry}, {Preparing Red-Green-Blue Images from CCD Data},
  PASP 116~(816) (2004) 133--137.
\newblock \href {http://arxiv.org/abs/astro-ph/0312483}
  {\path{arXiv:astro-ph/0312483}}, \href {http://dx.doi.org/10.1086/382245}
  {\path{doi:10.1086/382245}}.

\bibitem{Wang2018}
N.~{Wang}, J.~{Choi}, D.~{Brand}, C.-Y. {Chen}, K.~{Gopalakrishnan}, {Training
  Deep Neural Networks with 8-bit Floating Point Numbers}, arXiv e-prints
  (2018) arXiv:1812.08011\href {http://arxiv.org/abs/1812.08011}
  {\path{arXiv:1812.08011}}.

\bibitem{Ren2014}
W.~{Ren}, Y.~{Yu}, J.~{Zhang}, K.~{Huang}, Learning convolutional nonlinear
  features for k nearest neighbor image classification, in: 22nd International
  Conference on Pattern Recognition, 2014, pp. 4358--4363.

\bibitem{Honghui2016}
S.~{Honghui}, {Galaxy Classification with deep convolutional neural networks},
  Ph.D. thesis, University of Illinois at Urbana-Champaign (2016).

\bibitem{Meyer2018}
B.~J. {Meyer}, B.~{Harwood}, T.~{Drummond}, Deep metric learning and image
  classification with nearest neighbour gaussian kernels, in: 25th IEEE
  International Conference on Image Processing (ICIP), 2018, pp. 151--155.

\bibitem{Pan2020}
J.~Pan, V.~Pham, M.~Dorairaj, H.~Chen, J.-Y. Lee, Adversarial validation
  approach to concept drift problem in user targeting automation systems at
  uber (2020).
\newblock \href {http://arxiv.org/abs/2004.03045} {\path{arXiv:2004.03045}}.

\bibitem{Bishop1995}
C.~Bishop, {Neural networks for pattern recognition}, Oxford University Press,
  USA, 1995.

\bibitem{Dieleman2015}
S.~{Dieleman}, K.~W. {Willett}, J.~{Dambre}, {Rotation-invariant convolutional
  neural networks for galaxy morphology prediction}, MNRAS 450~(2) (2015)
  1441--1459.
\newblock \href {http://arxiv.org/abs/1503.07077} {\path{arXiv:1503.07077}},
  \href {http://dx.doi.org/10.1093/mnras/stv632}
  {\path{doi:10.1093/mnras/stv632}}.

\bibitem{he2015deep}
K.~He, X.~Zhang, S.~Ren, J.~Sun, Deep residual learning for image recognition
  (2015).
\newblock \href {http://arxiv.org/abs/1512.03385} {\path{arXiv:1512.03385}}.

\bibitem{Vega2021}
J.~{Vega-Ferrero}, H.~{Dominguez Sanchez}, M.~{Bernardi}, M.~e.~a.
  {Huertas-Company}, {Pushing automated morphological classifications to their
  limits with the Dark Energy Survey}, MNRAS 506~(2) (2021) 1927--1943.
\newblock \href {http://arxiv.org/abs/2012.07858} {\path{arXiv:2012.07858}},
  \href {http://dx.doi.org/10.1093/mnras/stab594}
  {\path{doi:10.1093/mnras/stab594}}.

\bibitem{Bhambra2022}
P.~{Bhambra}, B.~{Joachimi}, O.~{Lahav}, {Explaining deep learning of galaxy
  morphology with saliency mapping}, MNRAS 511~(4) (2022) 5032--5041.
\newblock \href {http://arxiv.org/abs/2110.08288} {\path{arXiv:2110.08288}},
  \href {http://dx.doi.org/10.1093/mnras/stac368}
  {\path{doi:10.1093/mnras/stac368}}.

\bibitem{Gupta2022}
R.~{Gupta}, P.~K. {Srijith}, S.~{Desai}, {Galaxy morphology classification
  using neural ordinary differential equations}, Astronomy and Computing 38
  (2022) 100543.
\newblock \href {http://arxiv.org/abs/2012.07735} {\path{arXiv:2012.07735}},
  \href {http://dx.doi.org/10.1016/j.ascom.2021.100543}
  {\path{doi:10.1016/j.ascom.2021.100543}}.

\bibitem{huang2018densely}
G.~Huang, Z.~Liu, L.~van~der Maaten, K.~Q. Weinberger, Densely connected
  convolutional networks (2018).
\newblock \href {http://arxiv.org/abs/1608.06993} {\path{arXiv:1608.06993}}.

\bibitem{szegedy2015rethinking}
C.~Szegedy, V.~Vanhoucke, S.~Ioffe, J.~Shlens, Z.~Wojna, Rethinking the
  inception architecture for computer vision (2015).
\newblock \href {http://arxiv.org/abs/1512.00567} {\path{arXiv:1512.00567}}.

\bibitem{szegedy2016inceptionv4}
C.~Szegedy, S.~Ioffe, V.~Vanhoucke, A.~Alemi, Inception-v4, inception-resnet
  and the impact of residual connections on learning (2016).
\newblock \href {http://arxiv.org/abs/1602.07261} {\path{arXiv:1602.07261}}.

\bibitem{Zoph2017}
B.~{Zoph}, V.~{Vasudevan}, J.~{Shlens}, {Learning Transferable Architectures
  for Scalable Image Recognition}, arXiv e-prints (2017) arXiv:1707.07012\href
  {http://arxiv.org/abs/1707.07012} {\path{arXiv:1707.07012}}.

\bibitem{simonyan2015deep}
K.~Simonyan, A.~Zisserman, Very deep convolutional networks for large-scale
  image recognition (2015).
\newblock \href {http://arxiv.org/abs/1409.1556} {\path{arXiv:1409.1556}}.

\bibitem{chollet2017xception}
F.~Chollet, Xception: Deep learning with depthwise separable convolutions
  (2017).
\newblock \href {http://arxiv.org/abs/1610.02357} {\path{arXiv:1610.02357}}.

\bibitem{Bradley1997}
A.~P. {Bradley}, {The use of the area under the ROC curve in the evaluation of
  machine learning algorithms}, Pattern Recognition 30~(7) (1997) 1145--1159.
\newblock \href {http://dx.doi.org/10.1016/S0031-3203(96)00142-2}
  {\path{doi:10.1016/S0031-3203(96)00142-2}}.

\bibitem{Rahmani2018}
S.~{Rahmani}, H.~{Teimoorinia}, P.~{Barmby}, {Classifying galaxy spectra at 0.5
  < z < 1 with self-organizing maps}, MNRAS 478~(4) (2018) 4416--4432.
\newblock \href {http://arxiv.org/abs/1805.07845} {\path{arXiv:1805.07845}},
  \href {http://dx.doi.org/10.1093/mnras/sty1291}
  {\path{doi:10.1093/mnras/sty1291}}.

\bibitem{Curti2022}
M.~{Curti}, C.~{Hayden-Pawson}, R.~{Maiolino}, F.~{Belfiore}, F.~{Mannucci},
  A.~{Concas}, G.~{Cresci}, A.~{Marconi}, M.~{Cirasuolo}, {What drives the
  scatter of local star-forming galaxies in the BPT diagrams? A Machine
  Learning based analysis}, MNRAS 512~(3) (2022) 4136--4163.
\newblock \href {http://arxiv.org/abs/2110.11841} {\path{arXiv:2110.11841}},
  \href {http://dx.doi.org/10.1093/mnras/stac544}
  {\path{doi:10.1093/mnras/stac544}}.

\bibitem{Shi2015}
F.~{Shi}, Y.-Y. {Liu}, G.-L. {Sun}, P.-Y. {Li}, Y.-M. {Lei}, J.~{Wang}, {A
  support vector machine for spectral classification of emission-line galaxies
  from the Sloan Digital Sky Survey}, MNRAS 453~(1) (2015) 122--127.
\newblock \href {http://dx.doi.org/10.1093/mnras/stv1617}
  {\path{doi:10.1093/mnras/stv1617}}.

\bibitem{Tempel2011}
E.~{Tempel}, E.~{Saar}, L.~J. {Liivam{\"a}gi}, A.~{Tamm}, J.~{Einasto},
  M.~{Einasto}, V.~{M{\"u}ller}, {Galaxy morphology, luminosity, and
  environment in the SDSS DR7}, Astron. Astrophys. 529 (2011) A53.
\newblock \href {http://arxiv.org/abs/1012.1470} {\path{arXiv:1012.1470}},
  \href {http://dx.doi.org/10.1051/0004-6361/201016196}
  {\path{doi:10.1051/0004-6361/201016196}}.

\bibitem{Tojeiro2013}
R.~{Tojeiro}, K.~L. {Masters}, J.~{Richards}, W.~J. {Percival}, S.~P.
  {Bamford}, C.~{Maraston}, R.~C. {Nichol}, R.~{Skibba}, D.~{Thomas}, {The
  different star formation histories of blue and red spiral and elliptical
  galaxies}, MNRAS 432~(1) (2013) 359--373.
\newblock \href {http://arxiv.org/abs/1303.3551} {\path{arXiv:1303.3551}},
  \href {http://dx.doi.org/10.1093/mnras/stt484}
  {\path{doi:10.1093/mnras/stt484}}.

\bibitem{Vavilova2015}
I.~B. {Vavilova}, G.~Y. {Ivashchenko}, I.~V. {Babyk}, O.~{Sergijenko}, D.~V.
  {Dobrycheva}, O.~O. {Torbaniuk}, A.~A. {Vasylenko}, N.~G. {Pulatova}, {The
  astrocosmic databases for multi-wavelength and cosmological properties of
  extragalactic sources}, Kosmichna Nauka i Tekhnologiya 21~(3) (2015) 94--107.
\newblock \href {http://dx.doi.org/10.15407/knit2015.05.094}
  {\path{doi:10.15407/knit2015.05.094}}.

\bibitem{Guo2020}
R.~{Guo}, C.-N. {Hao}, X.~{Xia}, Y.~{Shi}, Y.~{Chen}, S.~{Li}, Q.~{Gu}, {Toward
  an Understanding of the Massive Red Spiral Galaxy Formation}, Astrophys.J.
  897~(2) (2020) 162.
\newblock \href {http://arxiv.org/abs/2006.05462} {\path{arXiv:2006.05462}},
  \href {http://dx.doi.org/10.3847/1538-4357/ab9b75}
  {\path{doi:10.3847/1538-4357/ab9b75}}.

\bibitem{Mezcua2014}
M.~{Mezcua}, A.~P. {Lobanov}, E.~{Mediavilla}, M.~{Karouzos}, {Photometric
  Decomposition of Mergers in Disk Galaxies}, ApJ 784~(1) (2014) 16.
\newblock \href {http://arxiv.org/abs/1401.5920} {\path{arXiv:1401.5920}},
  \href {http://dx.doi.org/10.1088/0004-637X/784/1/16}
  {\path{doi:10.1088/0004-637X/784/1/16}}.

\bibitem{Simmons2017}
B.~D. {Simmons}, C.~{Lintott}, K.~W. {Willett}, K.~L. {Masters}, {et al.},
  {Galaxy Zoo: quantitative visual morphological classifications for 48 000
  galaxies from CANDELS}, MNRAS 464~(4) (2017) 4420--4447.
\newblock \href {http://arxiv.org/abs/1610.03070} {\path{arXiv:1610.03070}},
  \href {http://dx.doi.org/10.1093/mnras/stw2587}
  {\path{doi:10.1093/mnras/stw2587}}.

\bibitem{Bottrell2019}
C.~{Bottrell}, M.~H. {Hani}, H.~{Teimoorinia}, S.~L. {Ellison}, L.~{Moreno},
  {et al.}, {Deep learning predictions of galaxy merger stage and the
  importance of observational realism}, MNRAS 490~(4) (2019) 5390--5413.
\newblock \href {http://arxiv.org/abs/1910.07031} {\path{arXiv:1910.07031}},
  \href {http://dx.doi.org/10.1093/mnras/stz2934}
  {\path{doi:10.1093/mnras/stz2934}}.

\bibitem{Pearson2019}
W.~J. {Pearson}, L.~{Wang}, J.~W. {Trayford}, C.~E. {Petrillo}, F.~F.~S. {van
  der Tak}, {Identifying galaxy mergers in observations and simulations with
  deep learning}, A\&A 626 (2019) A49.
\newblock \href {http://arxiv.org/abs/1902.10626} {\path{arXiv:1902.10626}},
  \href {http://dx.doi.org/10.1051/0004-6361/201935355}
  {\path{doi:10.1051/0004-6361/201935355}}.

\bibitem{Cabrera2018}
G.~{Cabrera-Vives}, C.~J. {Miller}, J.~{Schneider}, {Systematic Labeling Bias
  in Galaxy Morphologies}, Astron.J. 156~(6) (2018) 284.
\newblock \href {http://arxiv.org/abs/1811.03577} {\path{arXiv:1811.03577}},
  \href {http://dx.doi.org/10.3847/1538-3881/aae9f4}
  {\path{doi:10.3847/1538-3881/aae9f4}}.

\bibitem{Hart2016}
R.~E. {Hart}, S.~P. {Bamford}, K.~W. {Willett}, K.~L. {Masters},
  C.~{Cardamone}, C.~J. {Lintott}, R.~J. {Mackay}, R.~C. {Nichol}, C.~K.
  {Rosslowe}, B.~D. {Simmons}, R.~J. {Smethurst}, {Galaxy Zoo: comparing the
  demographics of spiral arm number and a new method for correcting redshift
  bias}, MNRAS 461~(4) (2016) 3663--3682.
\newblock \href {http://arxiv.org/abs/1607.01019} {\path{arXiv:1607.01019}},
  \href {http://dx.doi.org/10.1093/mnras/stw1588}
  {\path{doi:10.1093/mnras/stw1588}}.

\bibitem{Tarsitano2022}
F.~{Tarsitano}, C.~{Bruderer}, K.~{Schawinski}, W.~G. {Hartley}, {Image feature
  extraction and galaxy classification: a novel and efficient approach with
  automated machine learning}, MNRAS 511~(3) (2022) 3330--3338.
\newblock \href {http://arxiv.org/abs/2105.01070} {\path{arXiv:2105.01070}},
  \href {http://dx.doi.org/10.1093/mnras/stac233}
  {\path{doi:10.1093/mnras/stac233}}.

\bibitem{Gauthier2016}
A.~{Gauthier}, A.~{Jain}, E.~{Noordeh},
  \href{http://cs229.stanford.edu/proj2016/report/GauthierJainNoordeh-GalaxyMorphology-report.pdf}{{Galaxy
  Morphology Classification}}, e-proceedings (2016) 1--6.
\newline\urlprefix\url{http://cs229.stanford.edu/proj2016/report/GauthierJainNoordeh-GalaxyMorphology-report.pdf}

\bibitem{Barchi2020}
P.~H. {Barchi}, R.~R. {de Carvalho}, R.~R. {Rosa}, R.~A. {Sautter}, {et al.},
  {Machine and Deep Learning applied to galaxy morphology - A comparative
  study}, Astronomy and Computing 30 (2020) 100334.
\newblock \href {http://arxiv.org/abs/1901.07047} {\path{arXiv:1901.07047}},
  \href {http://dx.doi.org/10.1016/j.ascom.2019.100334}
  {\path{doi:10.1016/j.ascom.2019.100334}}.

\bibitem{Mittal2020}
A.~{Mittal}, A.~{Soorya}, P.~{Nagrath}, D.~J. {Hemanth}, {Data augmentation
  based morphological classification of galaxies using deep convolutional
  neural network}, Earth Sci. Inform. 13 (2020) 601--617.
\newblock \href {http://dx.doi.org/10.1007/s12145-019-00434-8}
  {\path{doi:10.1007/s12145-019-00434-8}}.

\bibitem{Sreejith2018}
S.~{Sreejith}, J.~{Pereverzyev}, Sergiy, L.~S. {Kelvin}, F.~R. {Marleau},
  M.~{Haltmeier}, J.~{Ebner}, J.~{Bland-Hawthorn}, S.~P. {Driver}, A.~W.
  {Graham}, B.~W. {Holwerda}, A.~M. {Hopkins}, J.~{Liske}, J.~{Loveday}, A.~J.
  {Moffett}, K.~A. {Pimbblet}, E.~N. {Taylor}, L.~{Wang}, A.~H. {Wright},
  {Galaxy And Mass Assembly: automatic morphological classification of galaxies
  using statistical learning}, MNRAS 474~(4) (2018) 5232--5258.
\newblock \href {http://arxiv.org/abs/1711.06125} {\path{arXiv:1711.06125}},
  \href {http://dx.doi.org/10.1093/mnras/stx2976}
  {\path{doi:10.1093/mnras/stx2976}}.

\bibitem{Ghosh2020}
A.~{Ghosh}, C.~M. {Urry}, Z.~{Wang}, K.~{Schawinski}, D.~{Turp}, M.~C.
  {Powell}, {Galaxy Morphology Network: A Convolutional Neural Network Used to
  Study Morphology and Quenching in {\ensuremath{\sim}}100,000 SDSS and
  {\ensuremath{\sim}}20,000 CANDELS Galaxies}, Astrophys.J. 895~(2) (2020) 112.
\newblock \href {http://arxiv.org/abs/2006.14639} {\path{arXiv:2006.14639}},
  \href {http://dx.doi.org/10.3847/1538-4357/ab8a47}
  {\path{doi:10.3847/1538-4357/ab8a47}}.

\bibitem{Walmsley2022}
M.~{Walmsley}, A.~M.~M. {Scaife}, C.~{Lintott}, M.~{Lochner}, V.~{Etsebeth},
  T.~{G{\'e}ron}, H.~{Dickinson}, L.~{Fortson}, S.~{Kruk}, K.~L. {Masters},
  K.~B. {Mantha}, B.~D. {Simmons}, {Practical galaxy morphology tools from deep
  supervised representation learning}, MNRAS 513~(2) (2022) 1581--1599.
\newblock \href {http://arxiv.org/abs/2110.12735} {\path{arXiv:2110.12735}},
  \href {http://dx.doi.org/10.1093/mnras/stac525}
  {\path{doi:10.1093/mnras/stac525}}.

\bibitem{Gauci2010}
A.~{Gauci}, K.~{Zarb Adami}, J.~{Abela}, {Machine Learning for Galaxy
  Morphology Classification}, arXiv e-prints (2010) arXiv:1005.0390\href
  {http://arxiv.org/abs/1005.0390} {\path{arXiv:1005.0390}}.

\bibitem{Dominguez2018}
H.~{Dom{\'\i}nguez S{\'a}nchez}, M.~{Huertas-Company}, M.~{Bernardi},
  D.~{Tuccillo}, J.~L. {Fischer}, {Improving galaxy morphologies for SDSS with
  Deep Learning}, MNRAS 476~(3) (2018) 3661--3676.
\newblock \href {http://arxiv.org/abs/1711.05744} {\path{arXiv:1711.05744}},
  \href {http://dx.doi.org/10.1093/mnras/sty338}
  {\path{doi:10.1093/mnras/sty338}}.

\bibitem{YaoYuLin2021}
J.~{Yao-Yu Lin}, S.-M. {Liao}, H.-J. {Huang}, W.-T. {Kuo}, O.~{Hsuan-Min Ou},
  {Galaxy Morphological Classification with Efficient Vision Transformer},
  arXiv e-prints (2021) arXiv:2110.01024\href {http://arxiv.org/abs/2110.01024}
  {\path{arXiv:2110.01024}}.

\bibitem{Zhu2019}
X.-P. {Zhu}, J.-M. {Dai}, C.-J. {Bian}, Y.~{Chen}, S.~{Chen}, C.~{Hu}, {Galaxy
  morphology classification with deep convolutional neural networks}, Ap\&SS
  364~(4) (2019) 55.
\newblock \href {http://dx.doi.org/10.1007/s10509-019-3540-1}
  {\path{doi:10.1007/s10509-019-3540-1}}.

\bibitem{Dhar2022}
S.~{Dhar}, L.~{Shamir}, {Systematic biases when using deep neural networks for
  annotating large catalogs of astronomical images}, Astronomy and Computing 38
  (2022) 100545.
\newblock \href {http://arxiv.org/abs/2201.03131} {\path{arXiv:2201.03131}},
  \href {http://dx.doi.org/10.1016/j.ascom.2022.100545}
  {\path{doi:10.1016/j.ascom.2022.100545}}.

\bibitem{Smethurst2022}
R.~J. {Smethurst}, K.~L. {Masters}, B.~D. {Simmons}, I.~L. {Garland},
  T.~{G{\'e}ron}, B.~{H{\"a}u{\ss}ler}, S.~{Kruk}, C.~J. {Lintott},
  D.~{O'Ryan}, M.~{Walmsley}, {Quantifying the poor purity and completeness of
  morphological samples selected by galaxy colour}, MNRAS 510~(3) (2022)
  4126--4133.
\newblock \href {http://arxiv.org/abs/2112.04507} {\path{arXiv:2112.04507}},
  \href {http://dx.doi.org/10.1093/mnras/stab3607}
  {\path{doi:10.1093/mnras/stab3607}}.

\bibitem{Kautsch2006}
S.~J. {Kautsch}, E.~K. {Grebel}, F.~D. {Barazza}, I.~{Gallagher}, J.~S., {A
  catalog of edge-on disk galaxies. From galaxies with a bulge to superthin
  galaxies}, A\&A 445~(2) (2006) 765--778.
\newblock \href {http://arxiv.org/abs/astro-ph/0509294}
  {\path{arXiv:astro-ph/0509294}}, \href
  {http://dx.doi.org/10.1051/0004-6361:20053981}
  {\path{doi:10.1051/0004-6361:20053981}}.

\bibitem{Bizyaev2014}
D.~V. {Bizyaev}, S.~J. {Kautsch}, A.~V. {Mosenkov}, V.~P. {Reshetnikov}, N.~Y.
  {Sotnikova}, N.~V. {Yablokova}, R.~W. {Hillyer}, {The Catalog of Edge-on Disk
  Galaxies from SDSS. I. The Catalog and the Structural Parameters of Stellar
  Disks}, ApJ 787~(1) (2014) 24.
\newblock \href {http://arxiv.org/abs/1404.3072} {\path{arXiv:1404.3072}},
  \href {http://dx.doi.org/10.1088/0004-637X/787/1/24}
  {\path{doi:10.1088/0004-637X/787/1/24}}.

\bibitem{Lima2021}
C.~{Lima-Dias}, A.~{Monachesi}, S.~{Torres-Flores}, A.~{Cortesi}, {et al.}, {An
  environmental dependence of the physical and structural properties in the
  Hydra cluster galaxies}, MNRAS 500~(1) (2021) 1323--1339.
\newblock \href {http://arxiv.org/abs/2010.15235} {\path{arXiv:2010.15235}},
  \href {http://dx.doi.org/10.1093/mnras/staa3326}
  {\path{doi:10.1093/mnras/staa3326}}.

\bibitem{Dominguez2019}
H.~{Dom{\'\i}nguez S{\'a}nchez}, M.~{Huertas-Company}, M.~{Bernardi},
  S.~{Kaviraj}, J.~L. {Fischer}, T.~M.~C. {Abbott}, F.~B. {Abdalla},
  J.~{Annis}, S.~{Avila}, D.~{Brooks}, E.~{Buckley-Geer}, A.~{Carnero Rosell},
  M.~{Carrasco Kind}, J.~{Carretero}, C.~E. {Cunha}, C.~B. {D'Andrea}, L.~N.
  {da Costa}, C.~{Davis}, J.~{De Vicente}, P.~{Doel}, A.~E. {Evrard},
  P.~{Fosalba}, J.~{Frieman}, J.~{Garc{\'\i}a-Bellido}, E.~{Gaztanaga}, D.~W.
  {Gerdes}, D.~{Gruen}, R.~A. {Gruendl}, J.~{Gschwend}, G.~{Gutierrez}, W.~G.
  {Hartley}, D.~L. {Hollowood}, K.~{Honscheid}, B.~{Hoyle}, D.~J. {James},
  K.~{Kuehn}, N.~{Kuropatkin}, O.~{Lahav}, M.~A.~G. {Maia}, M.~{March},
  P.~{Melchior}, F.~{Menanteau}, R.~{Miquel}, B.~{Nord}, A.~A. {Plazas},
  E.~{Sanchez}, V.~{Scarpine}, R.~{Schindler}, M.~{Schubnell}, M.~{Smith},
  R.~C. {Smith}, M.~{Soares-Santos}, F.~{Sobreira}, E.~{Suchyta}, M.~E.~C.
  {Swanson}, G.~{Tarle}, D.~{Thomas}, A.~R. {Walker}, J.~{Zuntz}, {Transfer
  learning for galaxy morphology from one survey to another}, MNRAS 484~(1)
  (2019) 93--100.
\newblock \href {http://arxiv.org/abs/1807.00807} {\path{arXiv:1807.00807}},
  \href {http://dx.doi.org/10.1093/mnras/sty3497}
  {\path{doi:10.1093/mnras/sty3497}}.

\bibitem{Du2019}
W.~{Du}, C.~{Cheng}, H.~{Wu}, M.~{Zhu}, Y.~{Wang}, {Low Surface Brightness
  Galaxy catalogue selected from the {\ensuremath{\alpha}}.40-SDSS DR7 Survey
  and Tully-Fisher relation}, MNRAS 483~(2) (2019) 1754--1795.
\newblock \href {http://arxiv.org/abs/1811.04569} {\path{arXiv:1811.04569}},
  \href {http://dx.doi.org/10.1093/mnras/sty2976}
  {\path{doi:10.1093/mnras/sty2976}}.

\bibitem{Lingard2020}
T.~K. {Lingard}, K.~L. {Masters}, C.~{Krawczyk}, C.~{Lintott}, {et al.},
  {Galaxy Zoo Builder: Four-component Photometric Decomposition of Spiral
  Galaxies Guided by Citizen Science}, ApJ 900~(2) (2020) 178.
\newblock \href {http://arxiv.org/abs/2006.10450} {\path{arXiv:2006.10450}},
  \href {http://dx.doi.org/10.3847/1538-4357/ab9d83}
  {\path{doi:10.3847/1538-4357/ab9d83}}.

\bibitem{Schawinski2014}
K.~{Schawinski}, C.~M. {Urry}, B.~D. {Simmons}, L.~{Fortson}, {et al.}, {The
  green valley is a red herring: Galaxy Zoo reveals two evolutionary pathways
  towards quenching of star formation in early- and late-type galaxies}, MNRAS
  440~(1) (2014) 889--907.
\newblock \href {http://arxiv.org/abs/1402.4814} {\path{arXiv:1402.4814}},
  \href {http://dx.doi.org/10.1093/mnras/stu327}
  {\path{doi:10.1093/mnras/stu327}}.

\bibitem{Smirnov2022}
D.~V. {Smirnov}, V.~P. {Reshetnikov}, {The luminosity function of ringed
  galaxies}, arXiv e-prints (2022) arXiv:2209.06875\href
  {http://arxiv.org/abs/2209.06875} {\path{arXiv:2209.06875}}.

\bibitem{Hoyle2011}
B.~{Hoyle}, K.~L. {Masters}, R.~C. {Nichol}, E.~M. {Edmondson}, A.~M. {Smith},
  C.~{Lintott}, R.~{Scranton}, S.~{Bamford}, K.~{Schawinski}, D.~{Thomas},
  {Galaxy Zoo: bar lengths in local disc galaxies}, MNRAS 415~(4) (2011)
  3627--3640.
\newblock \href {http://arxiv.org/abs/1104.5394} {\path{arXiv:1104.5394}},
  \href {http://dx.doi.org/10.1111/j.1365-2966.2011.18979.x}
  {\path{doi:10.1111/j.1365-2966.2011.18979.x}}.

\bibitem{Reza2021}
M.~{Reza}, {Galaxy morphology classification using automated machine learning},
  Astronomy and Computing 37 (2021) 100492.
\newblock \href {http://dx.doi.org/10.1016/j.ascom.2021.100492}
  {\path{doi:10.1016/j.ascom.2021.100492}}.

\bibitem{Ahn2012}
C.~P. {Ahn}, R.~{Alexandroff}, C.~{Allende Prieto}, {et al.}, {The Ninth Data
  Release of the Sloan Digital Sky Survey: First Spectroscopic Data from the
  SDSS-III Baryon Oscillation Spectroscopic Survey}, ApJS 203~(2) (2012) 21.
\newblock \href {http://arxiv.org/abs/1207.7137} {\path{arXiv:1207.7137}},
  \href {http://dx.doi.org/10.1088/0067-0049/203/2/21}
  {\path{doi:10.1088/0067-0049/203/2/21}}.

\bibitem{Blanton2017}
M.~R. {Blanton}, M.~A. {Bershady}, B.~{Abolfathi}, F.~D. {Albareti},
  C.~{Allende Prieto}, A.~{Almeida}, J.~{Alonso-Garc{\'{\i}}a}, F.~{Anders},
  S.~F. {Anderson}, B.~{Andrews}, et~al., {Sloan Digital Sky Survey IV: Mapping
  the Milky Way, Nearby Galaxies, and the Distant Universe}, AJ 154 (2017) 28.
\newblock \href {http://arxiv.org/abs/1703.00052} {\path{arXiv:1703.00052}},
  \href {http://dx.doi.org/10.3847/1538-3881/aa7567}
  {\path{doi:10.3847/1538-3881/aa7567}}.

\bibitem{Wenger2000}
M.~{Wenger}, F.~{Ochsenbein}, D.~{Egret}, P.~{Dubois}, F.~{Bonnarel},
  S.~{Borde}, F.~{Genova}, G.~{Jasniewicz}, S.~{Lalo{\"e}}, S.~{Lesteven},
  R.~{Monier}, {The SIMBAD astronomical database. The CDS reference database
  for astronomical objects}, Astron. Astrophys. Supl. 143 (2000) 9--22.
\newblock \href {http://arxiv.org/abs/astro-ph/0002110}
  {\path{arXiv:astro-ph/0002110}}, \href
  {http://dx.doi.org/10.1051/aas:2000332} {\path{doi:10.1051/aas:2000332}}.

\end{thebibliography}
%\bibliography{mybibfile}

\end{document}